\documentclass[aps,prd,twocolumn,showpacs,superscriptaddress,nofootinbib,floatfix]{revtex4-1}

\usepackage{amsmath}   
\usepackage{graphicx}  
\usepackage{xspace}
\usepackage{units}
\usepackage{placeins}
\usepackage{float}
\usepackage{verbatim}
\usepackage{afterpage}
\usepackage{hyperref}
\usepackage{color,soul}

\newcommand{\minerva}{MINERvA}
\newcommand{\qelike}{quasielastic-like}
\newcommand{\tune}{MnvGENIE v1}
\newcommand{\tunevT}{MnvGENIE v2}
\newcommand{\tuneplot}{MnvGENIE v1}
\newcommand{\pt}{$p_{T}$}
\newcommand{\pz}{$p_{||}$}
\newcommand{\xsq}{$\chi^2$}
\newcommand{\numu}{$\nu_\mu$}
\newcommand{\anumu}{$\overline{\nu}_\mu$}
\newcommand{\muon}{$\mu^-$}
\newcommand{\qsq}{$Q^{2}$}
\newcommand{\uboone}{MicroBooNE}

\newcommand{\sizecheck}{0} 
\newcommand{\PRDsupp}{1}   
\ifnum\PRDsupp=0
  
\else
  
\fi

\newcommand{\Rutgers}{Rutgers, The State University of New Jersey, Piscataway, New Jersey 08854, USA}

\newcommand{\Florida}{University of Florida, Department of Physics, Gainesville, FL 32611}

\newcommand{\CBPF}{Centro Brasileiro de Pesquisas F\'{i}sicas, Rua Dr. Xavier Sigaud 150, Urca, Rio de Janeiro, Rio de Janeiro, 22290-180, Brazil}
\newcommand{\PUCP}{Secci\'{o}n F\'{i}sica, Departamento de Ciencias, Pontificia Universidad Cat\'{o}lica del Per\'{u}, Apartado 1761, Lima, Per\'{u}}

\newcommand{\Pittsburgh}{Department of Physics and Astronomy, University of Pittsburgh, Pittsburgh, Pennsylvania 15260, USA}
\newcommand{\Guanajuato}{Campus Le\'{o}n y Campus Guanajuato, Universidad de Guanajuato, Lascurain de Retana No. 5, Colonia Centro, Guanajuato 36000, Guanajuato M\'{e}xico}

\newcommand{\Tufts}{Physics Department, Tufts University, Medford, Massachusetts 02155, USA}
\newcommand{\WM}{Department of Physics, College of William \& Mary, Williamsburg, Virginia 23187, USA}
\newcommand{\FNAL}{Fermi National Accelerator Laboratory, Batavia, Illinois 60510, USA}

\newcommand{\MCLA}{Massachusetts College of Liberal Arts, 375 Church Street, North Adams, MA 01247}
\newcommand{\UMD}{Department of Physics, University of Minnesota -- Duluth, Duluth, Minnesota 55812, USA}

\newcommand{\UNI}{Universidad Nacional de Ingenier\'{i}a, Apartado 31139, Lima, Per\'{u}}
\newcommand{\Rochester}{University of Rochester, Rochester, New York 14627 USA}

\newcommand{\USM}{Departamento de F\'{i}sica, Universidad T\'{e}cnica Federico Santa Mar\'{i}a, Avenida Espa\~{n}a 1680 Casilla 110-V, Valpara\'{i}so, Chile}
\newcommand{\Geneva}{University of Geneva, 1211 Geneva 4, Switzerland}

\newcommand{\OregonState}{Department of Physics, Oregon State University, Corvallis, Oregon 97331, USA}
\newcommand{\oxford}{Oxford University, Department of Physics, Oxford, OX1 3PJ United Kingdom}

\newcommand{\upenn}{Department of Physics and Astronomy, University of Pennsylvania, Philadelphia, PA 19104}
\newcommand{\AMU}{AMU Campus, Aligarh, Uttar Pradesh 202001, India}

\newcommand{\Mohali}{Department of Physical Sciences, IISER Mohali, Knowledge City, SAS Nagar, Mohali - 140306, Punjab, India}

\newcommand{\york}{York University, Department of Physics and Astronomy, Toronto, Ontario, M3J 1P3 Canada}
\newcommand{\mateusfcarneiroThanks}{Now at Brookhaven National Laboratory}

\begin{document}

\title{Double-Differential Inclusive Charged-Current $\nu_\mu$ Cross Sections on Hydrocarbon in MINERvA at $\langle E_{\nu} \rangle \sim$ \unit[3.5]{GeV}}

\author{A.~Filkins}                       \affiliation{\WM}
\author{D.~Ruterbories}                   \affiliation{\Rochester}
\author{Y.~Liu}							  \affiliation{\WM}	


\author{Z.~Ahmad~Dar}                    \affiliation{\AMU}
\author{F.~Akbar}                         \affiliation{\AMU}
\author{O.~Altinok}                       \affiliation{\Tufts}
\author{D.A.~Andrade}                     \affiliation{\Guanajuato}
\author{M.~V.~Ascencio}                   \affiliation{\PUCP}
\author{A.~Bashyal}                       \affiliation{\OregonState}
\author{A.~Bercellie}                     \affiliation{\Rochester}
\author{M.~Betancourt}                    \affiliation{\FNAL}
\author{A.~Bodek}                         \affiliation{\Rochester}
\author{J.~L.~Bonilla}                    \affiliation{\Guanajuato}
\author{A.~Bravar}                        \affiliation{\Geneva}
\author{H.~Budd}                          \affiliation{\Rochester}
\author{G.~Caceres}                       \affiliation{\CBPF}
\author{T.~Cai}                           \affiliation{\Rochester}
\author{M.F.~Carneiro}\thanks{\mateusfcarneiroThanks}  \affiliation{\OregonState}  \affiliation{\CBPF}
\author{H.~da~Motta}                      \affiliation{\CBPF}
\author{S.A.~Dytman}                      \affiliation{\Pittsburgh}
\author{G.A.~D\'{i}az~}                   \affiliation{\Rochester}  \affiliation{\PUCP}
\author{J.~Felix}                         \affiliation{\Guanajuato}
\author{L.~Fields}                        \affiliation{\FNAL}
\author{R.~Fine}                          \affiliation{\Rochester}
\author{A.M.~Gago}                        \affiliation{\PUCP}
\author{H.~Gallagher}                     \affiliation{\Tufts}
\author{A.~Ghosh}                         \affiliation{\USM}  \affiliation{\CBPF}
\author{R.~Gran}                          \affiliation{\UMD}
\author{D.A.~Harris}                      \affiliation{\york}  \affiliation{\FNAL}
\author{S.~Henry}                         \affiliation{\Rochester}
\author{S.~Jena}                          \affiliation{\Mohali}
\author{D.~Jena}                          \affiliation{\FNAL}
\author{J.~Kleykamp}                      \affiliation{\Rochester}
\author{M.~Kordosky}                      \affiliation{\WM}
\author{D.~Last}                          \affiliation{\upenn}
\author{T.~Le}                            \affiliation{\Tufts}  \affiliation{\Rutgers}
\author{J.~LeClerc}                       \affiliation{\Florida}
\author{A.~Lozano}                        \affiliation{\CBPF}
\author{X.-G.~Lu}                         \affiliation{\oxford}
\author{E.~Maher}                         \affiliation{\MCLA}
\author{S.~Manly}                         \affiliation{\Rochester}
\author{W.A.~Mann}                        \affiliation{\Tufts}
\author{C.~Mauger}                        \affiliation{\upenn}
\author{K.S.~McFarland}                   \affiliation{\Rochester}
\author{A.M.~McGowan}                     \affiliation{\Rochester}
\author{B.~Messerly}                      \affiliation{\Pittsburgh}
\author{J.~Miller}                        \affiliation{\USM}
\author{J.G.~Morf\'{i}n}                  \affiliation{\FNAL}
\author{J.K.~Nelson}                      \affiliation{\WM}
\author{C.~Nguyen}                        \affiliation{\Florida}
\author{A.~Norrick}                       \affiliation{\WM}
\author{A.~Olivier}                       \affiliation{\Rochester}
\author{V.~Paolone}                       \affiliation{\Pittsburgh}
\author{G.N.~Perdue}                      \affiliation{\FNAL}  \affiliation{\Rochester}
\author{M.A.~Ram\'{i}rez}                 \affiliation{\Guanajuato}
\author{R.D.~Ransome}                     \affiliation{\Rutgers}
\author{H.~Ray}                           \affiliation{\Florida}
\author{H.~Schellman}                     \affiliation{\OregonState}
\author{C.J.~Solano~Salinas}              \affiliation{\UNI}
\author{H.~Su}                            \affiliation{\Pittsburgh}
\author{M.~Sultana}                       \affiliation{\Rochester}
\author{V.S.~Syrotenko}                   \affiliation{\Tufts}
\author{E.~Valencia}                      \affiliation{\WM}  \affiliation{\Guanajuato}
\author{M.Wospakrik}                      \affiliation{\Florida}
\author{C.~Wret}                          \affiliation{\Rochester}
\author{B.~Yaeggy}                        \affiliation{\USM}
\author{L.~Zazueta}                       \affiliation{\WM}

\collaboration{The MINER$\nu$A Collaboration}\ \noaffiliation
\date{\today}

\begin{abstract}
\minerva~reports inclusive charged-current cross sections for muon neutrinos on hydrocarbon in the NuMI beamline. We measured the double-differential cross section in terms of the longitudinal and transverse muon momenta, as well as the single-differential cross sections in those variables. The data used in this analysis correspond to an exposure of $3.34\times10^{20}$ protons on target with a peak neutrino energy of approximately \unit[3.5]{GeV}.  Measurements are compared to the GENIE, NuWro and GiBUU neutrino cross-section predictions, as well as a version of GENIE modified to produce better agreement with prior exclusive \minerva~measurements. None of the models or variants were able to successfully reproduce the data across the entire phase space, which includes areas dominated by each interaction channel.
\end{abstract}

\ifnum\sizecheck=0  
\maketitle
\fi

\section{Introduction}
\label{sec:Intro}
Precision neutrino oscillation measurements rely on accurate nuclear interaction models to estimate certain systematic uncertainties which can be a significant component of the total systematic uncertainty~\cite{Abe:2017vif}\cite{Acero:2019ksn}. Estimation of the incident neutrino energy based on the final state particles relies on these models~\cite{Acero:2019ksn}\cite{Abe:2015awa}. 
Precise measurements of the inclusive CC neutrino cross section (including all interaction channels, with the only the presence of a charged lepton required) in the sub-GeV to multi-GeV regime of $E_\nu$ illuminate the interplay of quasielastic (QE) scattering, baryon resonance production (RES) and deep inelastic scattering (DIS). This interplay involves aspects of neutrino-nucleus scattering and nuclear modeling which are not well understood; hence, its exploration is of current interest.  Moreover, CC inclusive measurements provide stringent tests for neutrino generators and provide a basis for refinement of models that can ultimately reduce systematic uncertainties in oscillation experiments.

In this article we present an inclusive double-differential charged-current (CC) cross section, as well as two single-differential cross sections. There are three attributes of inclusive neutrino cross section measurements that enhance their utility to the neutrino physics community. Firstly, they have straightforward signal definitions that allow for direct comparison between experiments. Secondly, inclusive cross section measurements have high statistical precision and small background contamination.
Finally, inclusive cross section measurements provide the opportunity to look at the entirety of a single generator prediction at once, allowing for examination of the interplay of the various interaction channels often studied exclusively. 

The new cross sections are presented as functions of the transverse and longitudinal muon momenta. Muon momentum is a well-defined quantity that can be reconstructed to a high precision (in comparison with measurements of the final-state hadronic system) which makes interpretation in the true parameter space less sensitive to model assumptions. The double-differential nature of the measurement allows for some separation of different interaction channels, with quasielastic (QE) interactions, baryon resonance production (RES), and deep inelastic scattering (DIS) each dominating different regions of phase space.

In order to take advantage of the model sensitivity of inclusive and double-differential cross sections, we compare to predictions of three neutrino interaction models: GENIE~\cite{Andreopoulos:2009rq}, NuWro~\citep{Golan:2012wx}, and GiBUU~\citep{Buss:2011mx}. 
Modified versions of GENIE are also shown, including two altered to achieve better agreement with prior \minerva~analyses~\cite{Ruterbories:2018gub}. Details on the interaction models used for this analysis are discussed in Sec.\ref{subsec:generators}.

Prior inclusive measurements by \minerva~include \numu~and \anumu~total cross sections as a function of neutrino energy on scintillator and carbon using the ``low-$\nu$" method~\cite{Devan:2016rkm}\cite{Ren:2017xov}. 
Recent inclusive cross section measurements from T2K~\cite{Abe:2018uhf} and \uboone~\cite{Adams:2019iqc}~have similarly presented results as a double-differential cross section in muon variables. 
Both of these measurements were performed with a lower average neutrino beam energy than used for \minerva. There are some similarities in the \qsq~regions probed by quasielastic interactions in the three experiments, however, \minerva~has greater accesses to inelastic interaction channels. The \uboone~measurement also used a different target nucleus, argon.
Total cross section measurements presented as functions of neutrino energy have also been reported by NOMAD~\cite{Wu:2007ab} with most interactions on carbon, and by MINOS~\cite{Adamson:2009ju}, CCFR~\cite{Seligman:1997fe} and NuTeV~\cite{Tzanov:2005kr} with most interactions on iron, as well as CHORUS~\cite{KayisTopaksu:2008aa} with most interactions on lead.

This work builds on previous results from \minerva, and particularly benefits from a reduced flux uncertainty. A measurement of neutrino-electron scattering improved the knowledge of the absolute neutrino flux~\cite{Park:2015eqa}. Additionally, hadron production data and particle yield measurements were used to constrain the normalization and shape of the flux~\cite{Aliaga:2016oaz}. These measurements have reduced \minerva's average flux uncertainty to 7\%~\cite{Park:2015eqa}. 
\section{Experiment}
\minerva~is a fine-grained detector situated in the NuMI neutrino beamline at Fermilab. \minerva~consists of 208 active hexagonal planes made up of triangular plastic scintillator strips, with a region of nuclear targets (not used in this analysis), as well as an active tracking region~\cite{Aliaga:2013uqz}. This analysis uses a portion of the active tracking region with a fiducial mass of 5.48 tons. Electromagnetic and hadronic calorimeters surround the perimeter of these hexagonal planes, with additional electromagnetic and hadronic calorimetry downstream of the active tracking volume. The fiducial volume is comprised of 88.5\% carbon, 8.2\% hydrogen, 2.5\% oxygen, 0.5\% titanium, 0.2\% chlorine, 0.07\% aluminum, and 0.07\% silicon by mass. The strips in successive planes are arranged in three different orientations (0$^{\circ}$ and $\pm 60^{\circ}$ from vertical) to allow for three-dimensional track reconstruction. Wavelength-shifting fibers embedded in the strips of scintillator are read out by optical cables that connect to photomultiplier tubes. The photomultiplier tubes read out the scintillation light with a 3-ns timing resolution.

Muons exiting the downstream end of \minerva~may enter the MINOS near detector (ND), which sits 2 m downstream of \minerva. The MINOS ND is then able to measure muon charge and momentum~\cite{Michael:2008bc}.

The NuMI beam is produced by 120-GeV protons interacting with a carbon target. Magnetic focusing horns are used to direct positively charged mesons toward the \minerva~detector~\citep{Adamson:2015dkw}. The mesons decay in a helium-filled decay pipe producing a neutrino beam. The horizontal ($\hat{z}$) axis of the detector is at a 58-mrad angle relative to the direction of the NuMI beam, which points downward.

This analysis is based on data taken between 2010 and 2012 with an exposure of $3.34\times 10^{20}$ protons on target while the NuMI beam was operated in the low-energy neutrino mode. This mode provides a beam that is approximately 93\% muon neutrinos, 6\% muon antineutrinos, and 1\% electron neutrinos and antineutrinos, with a peak energy of approximately 3.5 GeV \cite{Aliaga:2016oaz}.

\section{Simulation}
\label{sec:Simulation}
\subsection{Detector response}
The simulation of detector response is based upon GEANT4 v4.9.4p6~\cite{Agostinelli2003250}; it includes an overlay of data events in order to model the effects of simultaneous activity taking place in the detector. A scaled down version of the \minerva~detector, which collected data in a charged-particle beam, was used to determine the absolute hadron energy scale and its uncertainty~\cite{Aliaga:2015aqe}. The response of the detector to minimum ionizing particles is calibrated using muons that transverse the length of the detector~\citep{Aliaga:2013uqz}. 

\subsection{Flux model}

The NuMI beam flux is modeled based on GEANT4~\citep{Adamson:2015dkw} with additional modifications derived from prior measurements of proton-carbon hadron production~\citep{Alt:2006fr}, as well as measured thin-target yields~\cite{Aliaga:2016oaz}. Neutrino-electron scattering, which was previously measured by \minerva~\cite{Park:2015eqa}, is also used to constrain the flux. 

\subsection{Interaction models}
\label{subsec:generators}
Neutrino interactions are simulated using GENIE 2.8.4~\cite{Andreopoulos:2009rq}. Nuclear effects are modeled 
using the relativistic Fermi gas model~\cite{Smith:1972xh} with a maximum momentum for a struck nucleon of $\unit[0.221]{GeV/c}$ and the Bodek-Ritchie short range correlation model for the inclusion of higher momentum struck nucleons~\cite{Bodek:1981wr}.

The Llewellyn-Smith formalism~\citep{LlewellynSmith:1971zm} with electromagnetic form factors from BBBA2005~\citep{Bradford:2006yz} is used for modeling quasielastic interactions. The axial form factor is assumed to have a dipole form and an axial vector mass of $M_A=0.99$ GeV/c$^2$. Resonance production in GENIE is simulated using the Rein-Seghal model~\citep{Rein:1980wg}. The Bodek-Yang model~\citep{Bodek:2004pc} is used to leading order for simulation of DIS. GENIE models hadron rescattering (final-state interactions) using the GENIE INTRANUKE-hA package~\cite{Dytman:2007zz}. 
In place of a full intranuclear cascade, final state interactions are modeled using an effective particle cascade. At most one particle rescatter is allowed before absorption or exiting the nucleus, with pion-nucleus scattering data used to determine the relative scattering probabilities\cite{Dytman:2011zz}.

\minerva~has made modifications to GENIE 2.8.4 in order to obtain better predictions of specific channels previously measured by \minerva, which is referred to as \tune~below.
A major modification was made to add a screening effect and its uncertainty to quasielastic reactions based on the Valencia group's random phase approximation (RPA) applied to a Fermi gas \cite{Nieves:2004wx,Gran:2017psn}. Another major modification was to add a meson exchange current based two-particle knockout process (leaving two holes in the nucleus, abbreviated 2p2h) \cite{Nieves:2011pp, Gran:2013kda, Schwehr:2016pvn}. These modifications did not provide sufficient strength to reproduce prior \minerva~ measurements of inclusive CC scattering at low momentum transfer \cite{Rodrigues:2015hik}. The 2p2h model was enhanced using an empirical fit to the observed hadronic energy spectrum achieving a good description of \cite{Rodrigues:2015hik}  by construction and the companion antineutrino data \cite{Gran:2018fxa} without further tuning. These modifications also improve the description of muon kinematics of CC events without pions \cite{Ruterbories:2018gub, Patrick:2018gvi} and the distribution of observed hadronic energy. Turning these modifications on and off is a major part of the discussion later in the paper. 

One more modification is made. The GENIE nonresonant pion production model (part of the GENIE DIS classification) is decreased by 43\% based on comparing GENIE to a reanalysis of deuterium bubble chamber data \cite{Wilkinson:2014yfa, Rodrigues:2016xjj}. This modification is made to all of the variations of GENIE that are used in this paper, with the exception of GENIE 2.8.4 which has no modifications.

A second version of the tune developed by \minerva, referred to as \minerva~GENIE v2, is also used as a model comparison.  \minerva~GENIE v2 includes all of the modifications used in \tune, with the addition of a suppression of pion production at low four-momentum squared ($Q^2$)~\cite{Stowell:2019zsh}. This suppression is tuned to prior \minerva~measurements of charged-current baryon resonance production that observed diminished event rates at low $Q^2$ ~\cite{Eberly:2014mra}\cite{McGivern:2016bwh}\cite{Altinok:2017xua}. 
A quantitatively similar suppression based on MINOS data is also included as a comparison~\cite{Adamson:2014pgc}.

NuWro~\citep{Golan:2012wx} and GiBUU~\citep{Buss:2011mx} simulations represent alternative interaction models that can be compared with these measurements. Additionally, three models of true deep inelastic scattering ($W>\unit[2.0]{GeV}$, $Q^2>\unit[1.0]{GeV^2}$) are shown as partial model comparisons. 
The \minerva~low-energy data used for this analysis spans a neutrino energy range from the first onset of nonresonant pion production to energies at which true DIS is the dominant CC interaction channel (all included in the GENIE DIS classification). Because DUNE will operate in a similar neutrino energy range, comparisons of the data with specific true DIS models are of particular interest. 
The first two true DIS models are nCTEQ15~\cite{Kovarik:2015cma} and nCTEQ$\nu$~\cite{Schienbein:2007fs}, which are global analyses of nuclear parton distributions based on charged lepton-nucleus and neutrino-nucleus scattering respectively. The third is a beyond leading order microscopic model developed at Aligarh Muslim University (AMU) referred to in plots as AMU DIS~\citep{amu:2016}. 
These true DIS models are implemented by reweighting GENIE DIS events which have $W$ and \qsq~values within the range for true DIS interactions mentioned above. The AMU and GENIE DIS models do not incorporate QED radiative corrections; however, these radiative corrections are included in the nCTEQ15 and nCTEQ$\nu$ fits. The other processes are described by \tune. 

\section{Event Sample}

\subsection{Signal Definition}
\begin{figure*}[p]
	\centering
	\includegraphics[width=\textwidth]{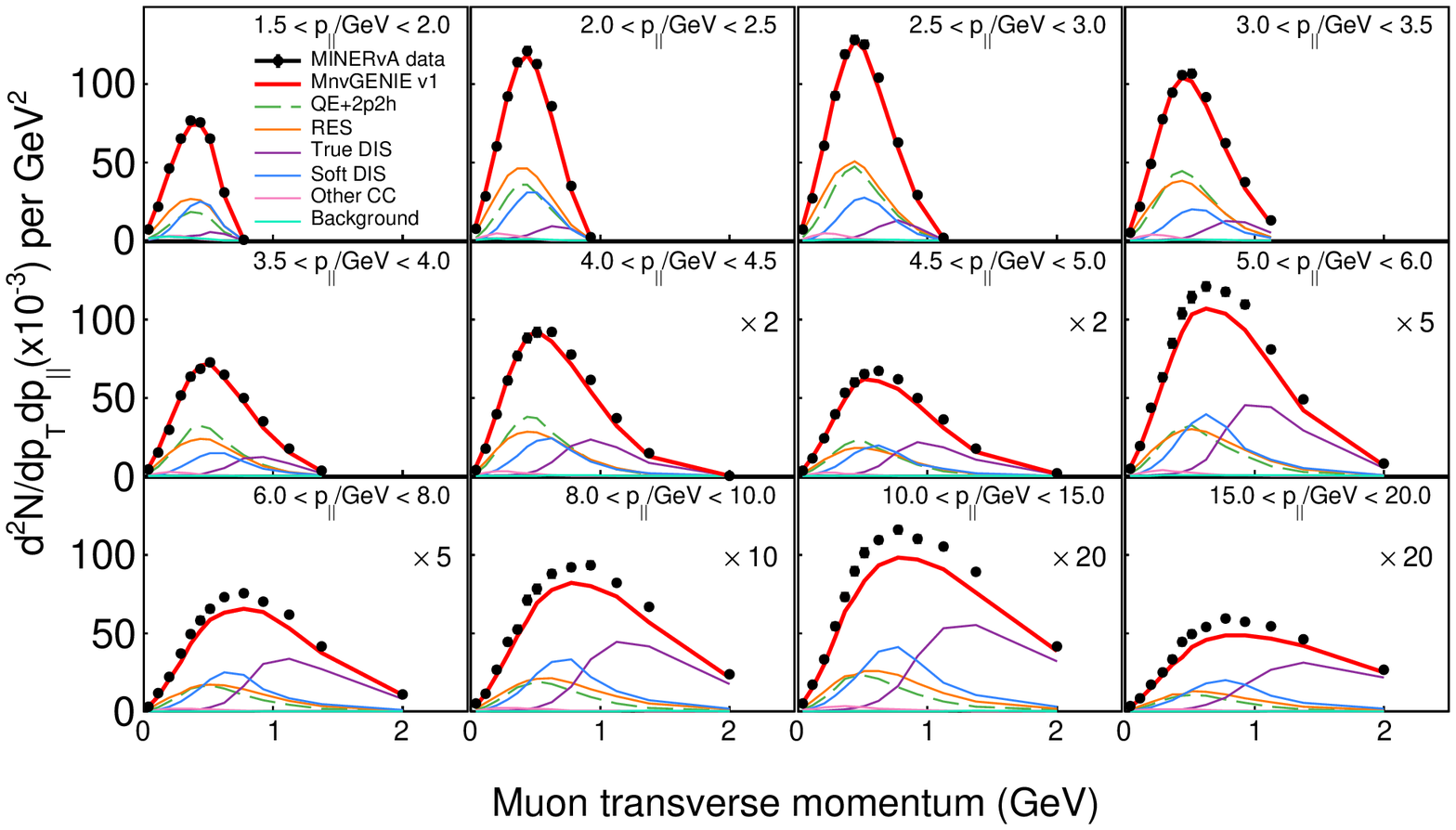} 
	\includegraphics[width=\textwidth]{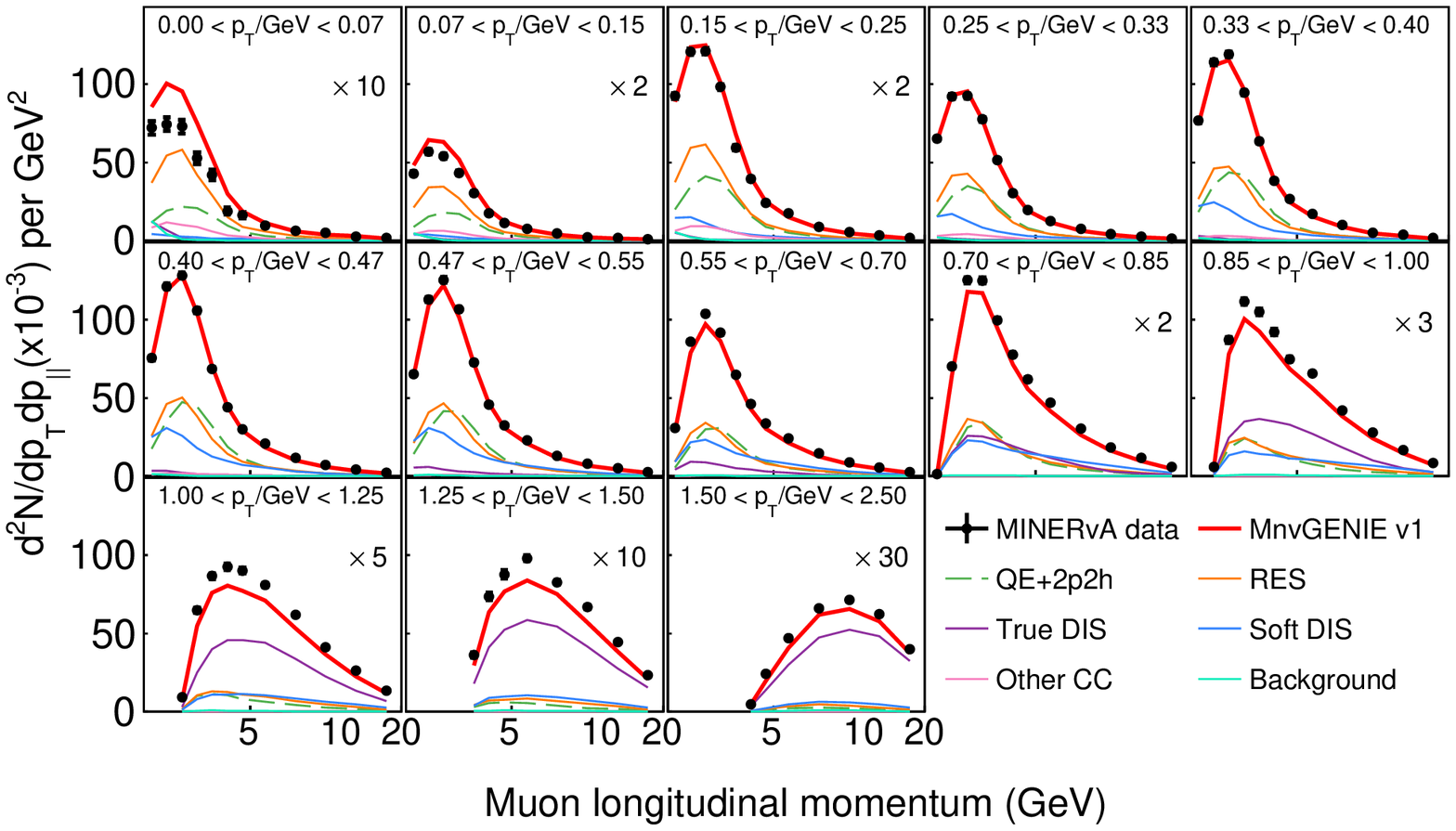} 
	\caption{
		Selected events of data and \tuneplot~in bins of longitudinal and transverse momenta, shown alongside modeled neutrino interaction types. Each panel represents a single bin of longitudinal (transverse) momentum in the top (bottom) plot. Note that multipliers are applied in some panels to better display low-population bins. }
	\label{fig:evsel_evtbymodelptpz}
\end{figure*}
The defining characteristic of a \numu~CC event is the presence of a \muon.  In order to be considered signal, an event must have one negatively-charged muon with an angle of less than 20 degrees with respect to the beamline.
An angular cut is needed because the acceptance of the MINOS near detector (which is required to reconstruct muon charge and momentum) decreases rapidly for events with muon scattering angles greater than 20 degrees.   There are no additional limitations made based on particle type, and events with any number of additional particles are allowed.

\subsection{Event Reconstruction and Selection}
\label{sec:eventreco}

The essential requirement for an event to be reconstructed in this analysis is that there is a muon present, and that the muon momentum, angle, and charge can all be reconstructed.
Muon track reconstruction requires the muon to originate in the fiducial volume, traverse the remainder of the \minerva~detector and leave a track in the MINOS near detector, which is matched with a \minerva~track. Timing and position information are used in order to match tracks in \minerva~with tracks in the MINOS near detector. 

This analysis reconstructs events using the same method described in Ref.~\citep{Ruterbories:2018gub}, though many of the variables reconstructed in the referenced exclusive \minerva~measurement are not used for this inclusive analysis, which only utilizes the muon. For a track to be reconstructed, the muon must traverse a minimum of 9 planes in \minerva.

The muon momentum is calculated by using the ionization energy loss for a muon traversing the material in the \minerva~detector in conjunction with the momentum reconstructed from MINOS~\cite{Michael:2008bc}. Muon charge is reconstructed using track curvature in the magnetized MINOS near detector and is required to be negative.

The reconstructed vertex must be within the fiducial volume of the active tracker area of the detector. A muon is classified as originating in the fiducial volume if its primary interaction vertex is located in a 2.37-m long section of scintillator 
and within an 850-mm apothem.

Tracks that are not associated with the primary event and activity occurring more then 5 ns before or 10 ns after the muon time are removed, as described in~\citep{Ruterbories:2018gub}.

\subsection{Selected Events}
\label{sec:eventselection}

The resulting event sample, after cuts but before background subtraction, is shown in Fig.~\ref{fig:evsel_evtbymodelptpz}.  This sample has 325,588 events with a selection purity (percentage of selected MC events that are true signal events) of 99.3\%.  In the figure, the unstacked components of \tune~are shown by GENIE interaction type. Quasielastic and two particle two hole (2p2h) events are combined into one category, while events classified as DIS by GENIE are further broken down using kinematic restrictions. True DIS events are defined as GENIE DIS events that have a $W>\unit[2.0]{GeV}$ and a $Q^2>\unit[1.0]{GeV^2}$, while, for this analysis, the soft DIS category is defined to contain the remaining GENIE DIS events that fail either or both of these kinematic limitations. Baryon single-pion resonance production and other CC events make up the remainder of the signal sample, while neutral-current events and charged-current events originating from other flavor neutrinos or any flavor of antineutrinos comprise the background events for this analysis. 

\section{Cross-section Extraction}
\label{sec:extraction}

In order to extract the cross section, we take the selected events and subtract the number of background events predicted by the simulation based on the total number of protons on target. The background-subtracted event sample is then unfolded in order to account for detector resolution effects. Next, an efficiency correction is applied to the unfolded sample. Finally, the efficiency-corrected sample is normalized by the flux and number of targets. 

\subsection{Background Subtraction}
In this inclusive sample, background events make up only 0.75\%  of the total selected simulated events with 0.50\% of the simulated events coming from other-flavor neutrino events (mostly  \anumu~events), and 0.24\% coming from neutral-current events (primarily from pion punch-through). 
The neutral-current events occur mostly in the lowest \pz~bin (1.5\,\textless\,\pz\,\textless\,\unit[2.0]{GeV}), with a peak in the third \pt~bin (0.15\,\textless\,\pt\,\textless\,\unit[0.25]{GeV}). 
The other-flavor neutrino events are more evenly distributed throughout the longitudinal momentum space, with a peak in the fifth \pt~bin (0.33\,\textless\,\pt\,\textless\,\unit[0.40]{GeV}) and first \pz~bin.

Background events make up less than 2 percent of selected events in over 95\% of the phase space, with larger background contributions appearing only in the lowest-momentum bins. The bin with the lowest total momentum has the largest contribution, with 15\% of events coming from background events. 
The small background contribution allows for an absolutely predicted background subtraction without introducing substantial model dependence.

\subsection{Unfolding}
\begin{figure}[tp]
	\centering
	\includegraphics[width=0.95\linewidth]{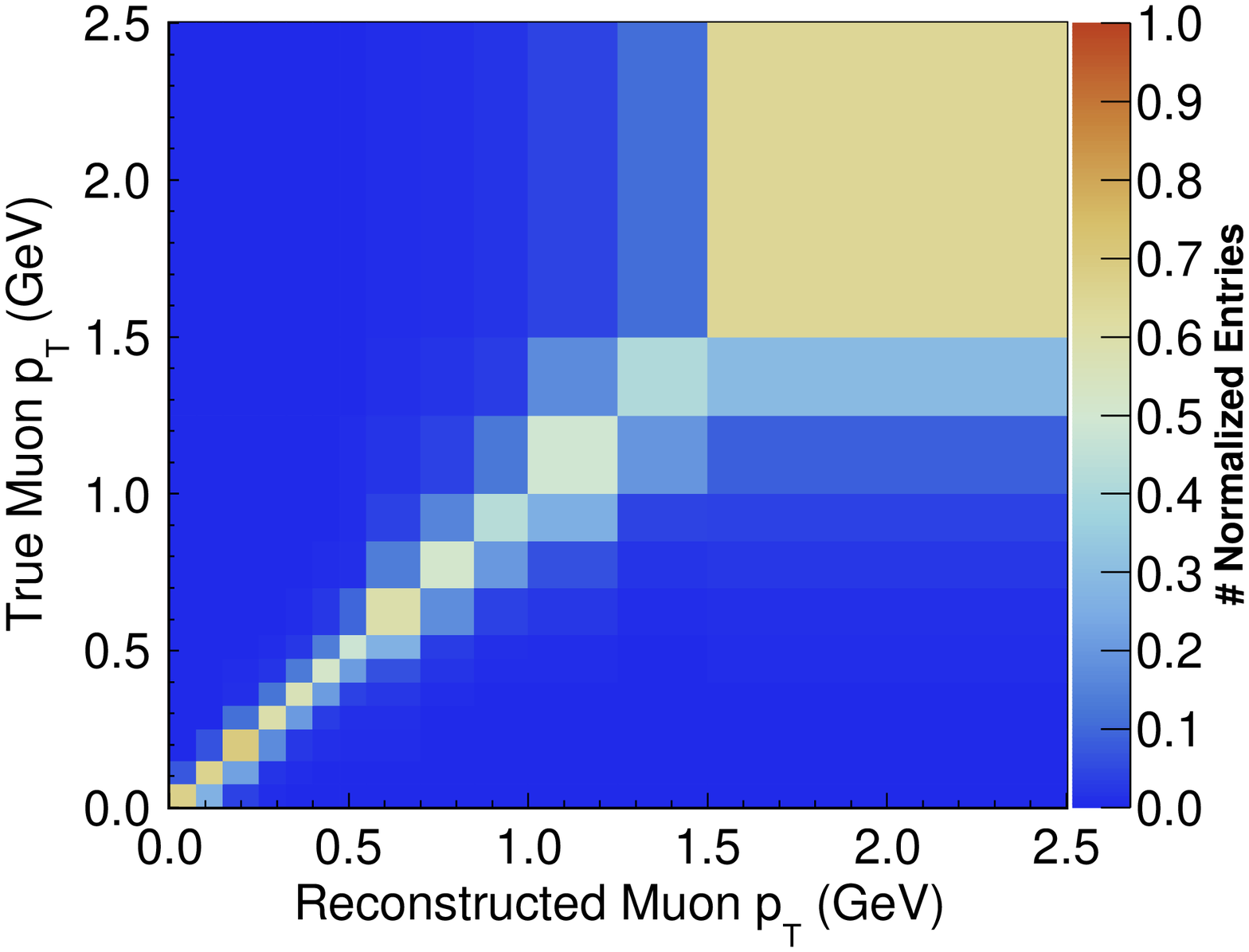} 
	\includegraphics[width=0.95\linewidth]{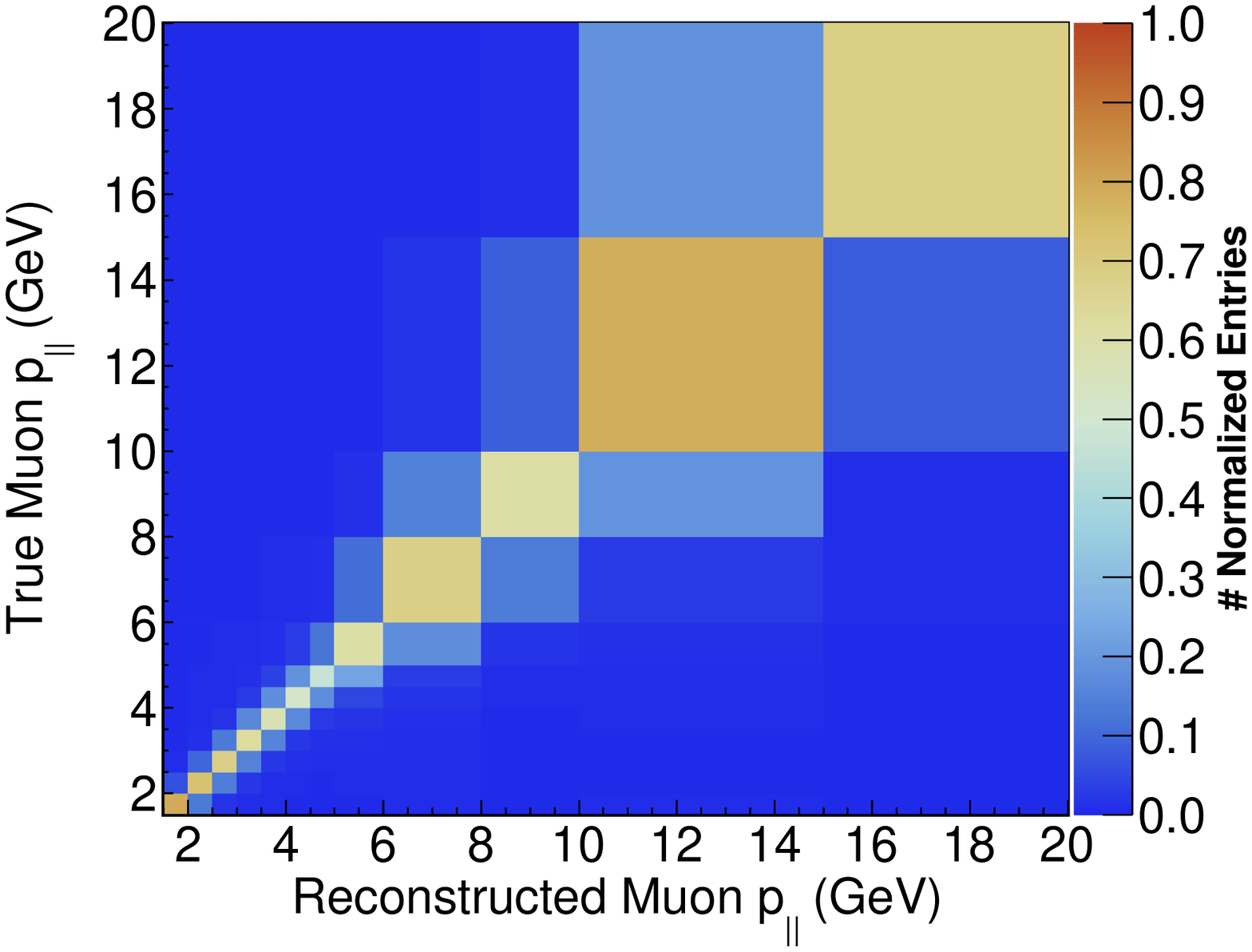} 
	\caption{Two-dimensional migration matrix projected into transverse momentum (top) and longitudinal momentum (bottom). Both projections are nearly diagonal.}
	\label{fig:mig}
\end{figure}
\begin{figure}[tp]
	\centering
	\includegraphics[width=0.95\linewidth]{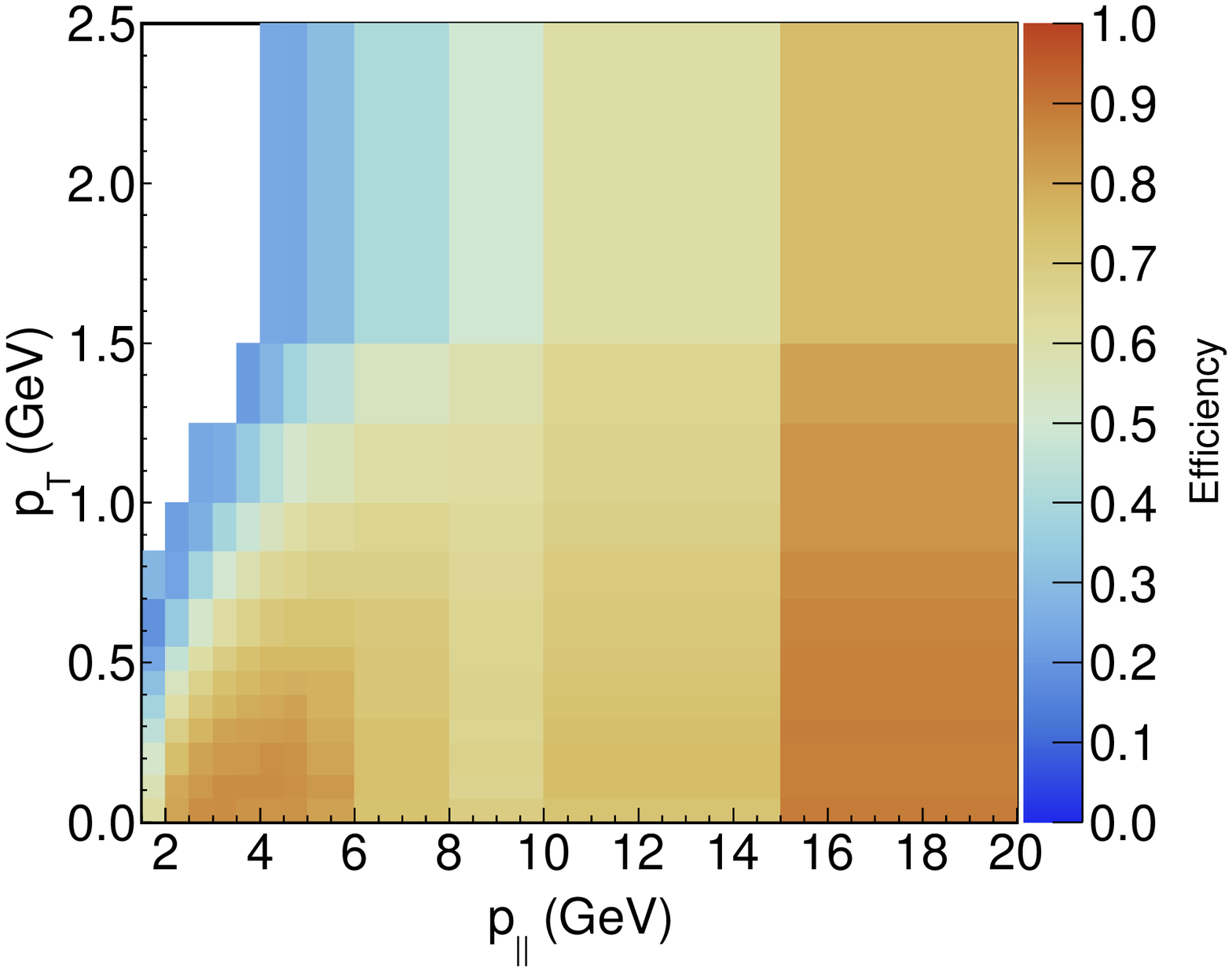} 
	\caption{Efficiency for the \numu~CC inclusive signal in bins of muon longitudinal and transverse momentum. The efficiency is highest (\textgreater\,85\%) for high \pz~and low \pt, corresponding to good acceptance into the MINOS ND.}
	\label{fig:eff_ptpz}
\end{figure}

\begin{figure}[tp]
	\centering
	\includegraphics[width=0.95\linewidth]{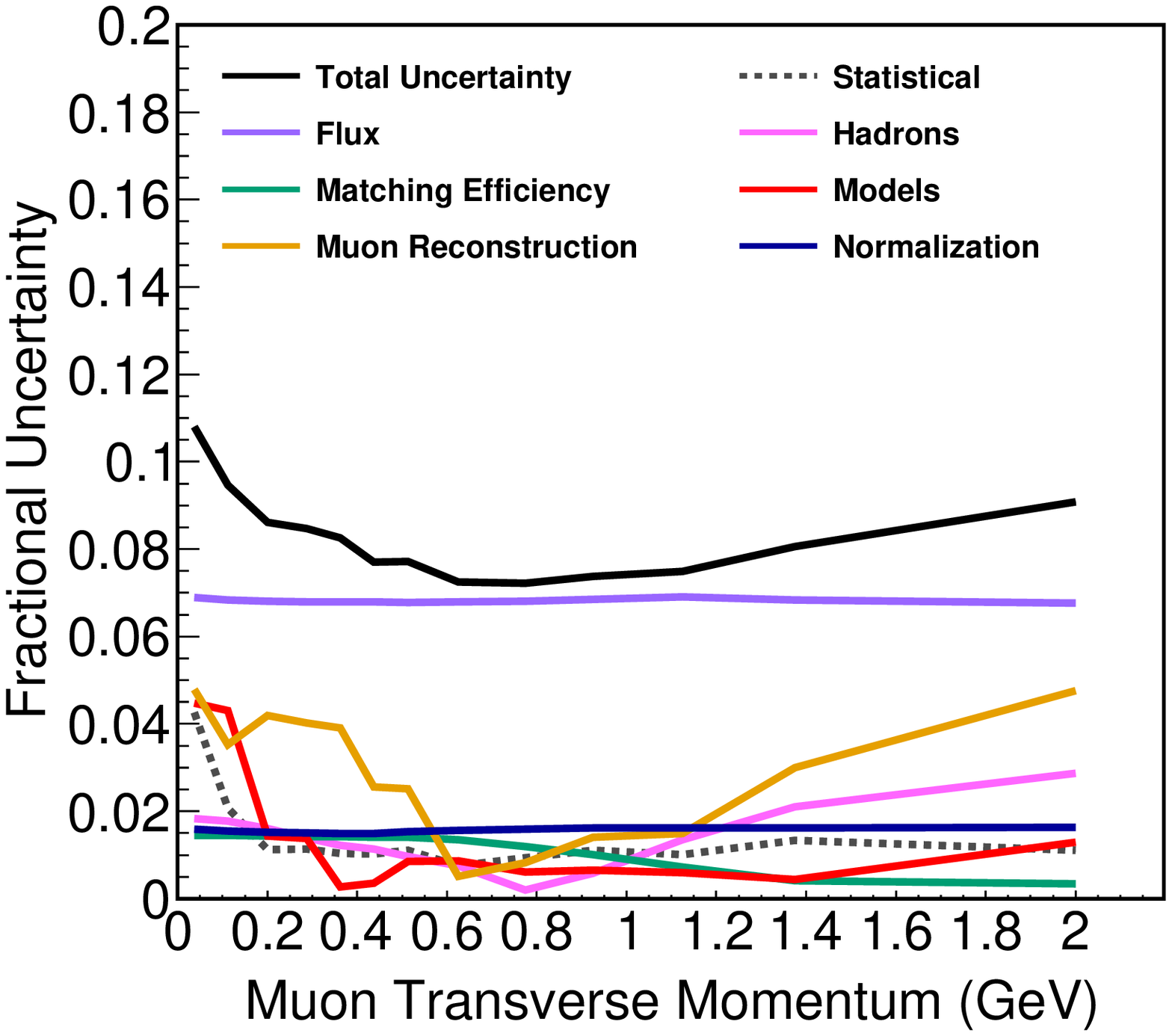} 
	\includegraphics[width=0.95\linewidth]{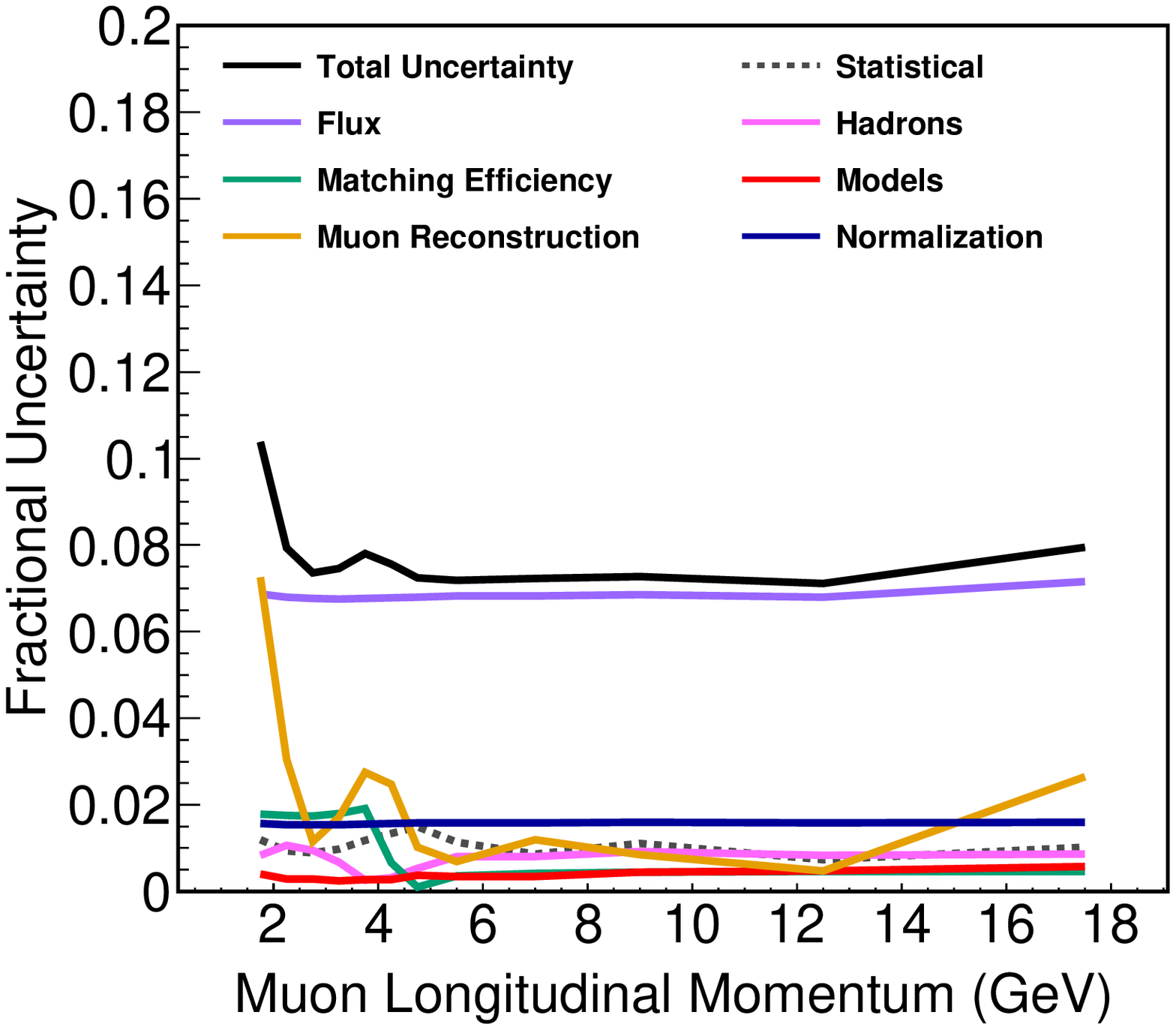} 
	\caption{Systematic uncertainties for \pt~(top) and \pz~(bottom) single-differential cross sections. The neutrino flux is the largest fractional uncertainty (7\%) for both variables, with the uncertainty from the muon reconstruction becoming sizable at high \pt~and low \pz.}
	\label{fig:1dsys}
\end{figure}

\begin{figure*}[tp]
	\centering
	\includegraphics[width=1\linewidth]{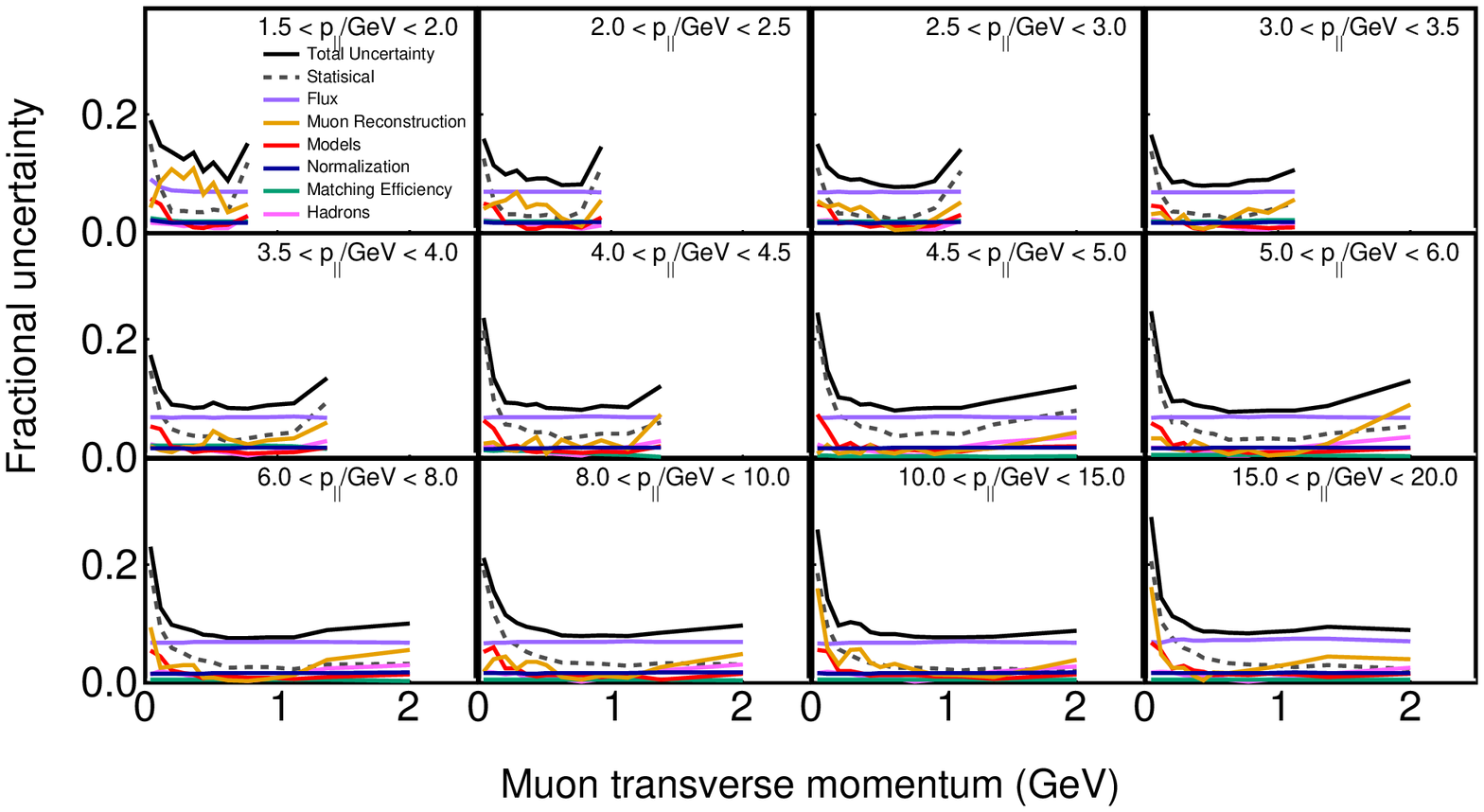} 
	\includegraphics[width=1\linewidth]{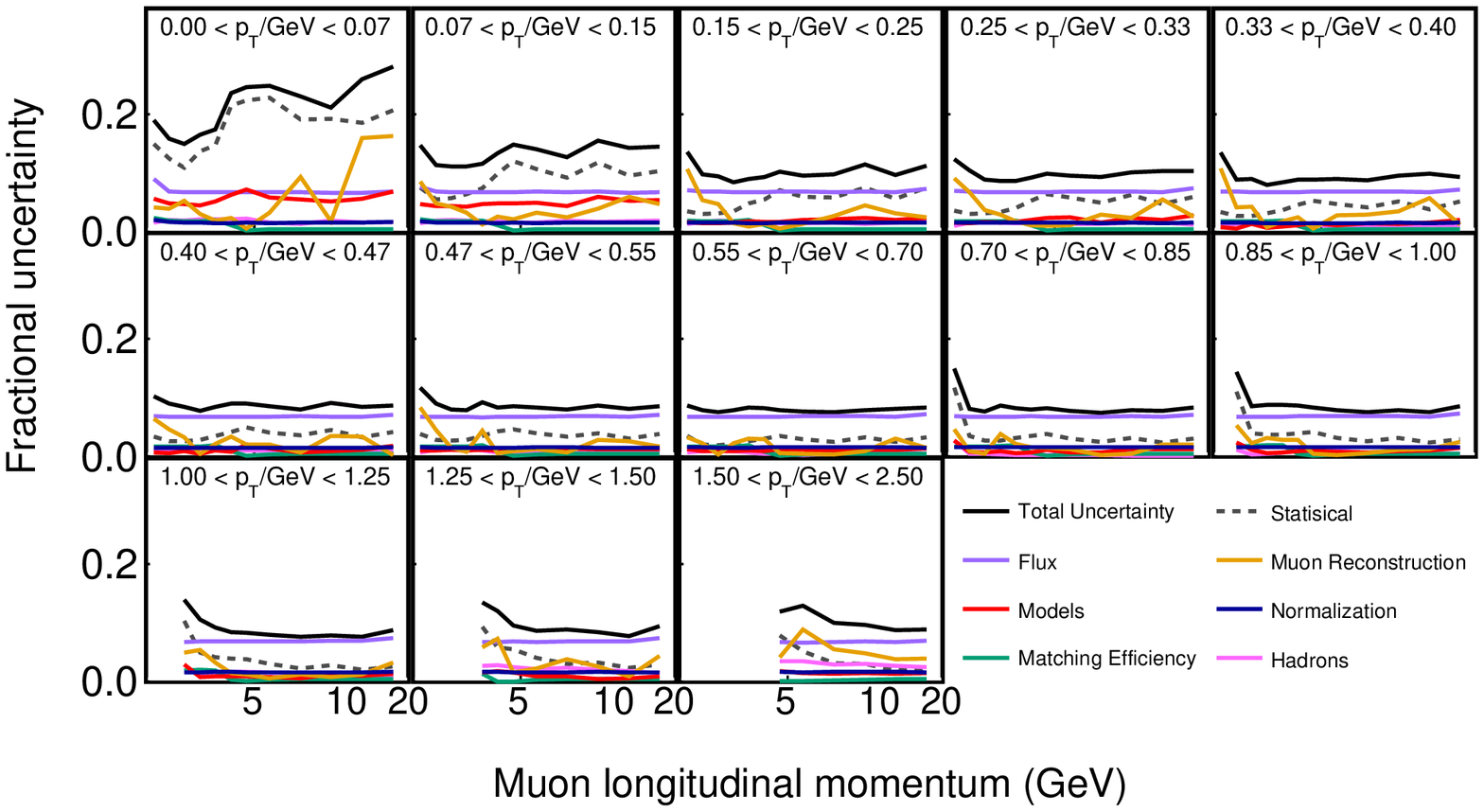} 
	\caption{Categorical breakdown of systematic uncertainties on the double-differential cross section measurement in slices of \pz~(top) and \pt~(bottom). Flux is the dominant uncertainty in the majority of phase space, with muon reconstruction the other comparable uncertainty in some areas.}
	
	\label{fig:sys_ptpz}
\end{figure*}
Detector resolution effects result in reconstructed variables being smeared away from their true values. The magnitude of the smearing can be estimated from Monte Carlo simulations of the detector response. The data are iteratively unfolded using RooUnfold~\cite{Adye:2011gm}, a ROOT implementation of the D'Agostini unfolding method~\citep{DAgostini:1994fjx}\citep{DAgostini:2010hil}. In order to estimate the validity of the unfolding method, several unfolding studies were done. 
For each unfolding study, a different reconstructed Monte Carlo event sample was used as an approximation of the response of the data (as pseudodata), and unfolded using the central value Monte Carlo smearing matrix. The number of iterations with which the pseudodata is unfolded is varied, and \xsq~values are calculated by comparing the unfolded pseudodata 
with event distributions of the pseudodata in true momentum space. In the first of these studies, we tested the unfolding by using the central value Monte Carlo for both the pseudodata and the smearing matrix. During this test, the $\chi^2$ reached the number of degrees of freedom within a single iteration as expected. Next, default GENIE 2.8.4 (no tunes applied), and GENIE with the nonresonant pion tune and quasielastic RPA applied (without the inclusion of a 2p2h sample), were used as pseudodata, while still using the smearing matrix derived from \tune. In these studies, the $\chi^2$ reached a minimum at 10 iterations. There are 144 degrees of freedom for these studies, as there are 12$\times$13 bins minus 12 bins excluded due to the 20 degree angle requirement. Additionally, a study was done in which the Monte Carlo was reweighted on an event by event basis (warped) using a weighting function that approximates the data to Monte Carlo ratio for \tune. The results using warped Monte Carlo as pseudodata were consistent with the former studies.  As a result of these studies, unfolding is performed with 10 iterations for this analysis.

The one-dimensional projections of the full two-dimensional migration matrix into \pt~ and \pz~are shown in Fig.~\ref{fig:mig}. The smearing matrices for both the full two-dimensional space and the projections are nearly diagonal.

\subsection{Efficiency Correction and Normalization}
\label{sec:Norm}

A bin-by-bin efficiency correction derived from the simulation is applied to the unfolded event sample; the signal efficiency is shown in Fig.~\ref{fig:eff_ptpz}. 

The empty region in the top left of the plot is due to the requirement that the muon must be within a 20 degree angle of the beamline. The area around this region has lower efficiencies due to a larger portion of the event muons missing MINOS at these larger angles. The efficiencies for each interaction type were also calculated, and they each have similar magnitudes and shapes to the total CC efficiency. 

Efficiency-corrected event rates are normalized using the flux given in~\citep{Aliaga:2016oaz} integrated from 0 to 120 GeV, resulting in a normalization factor of $2.877\times10^{-8}$cm$^{-2}$ per POT. Flux-averaged cross sections are then normalized by the number of nucleons in the fiducial volume. 
\begin{figure}[tp]
	\centering
	\includegraphics[width=0.95\linewidth]{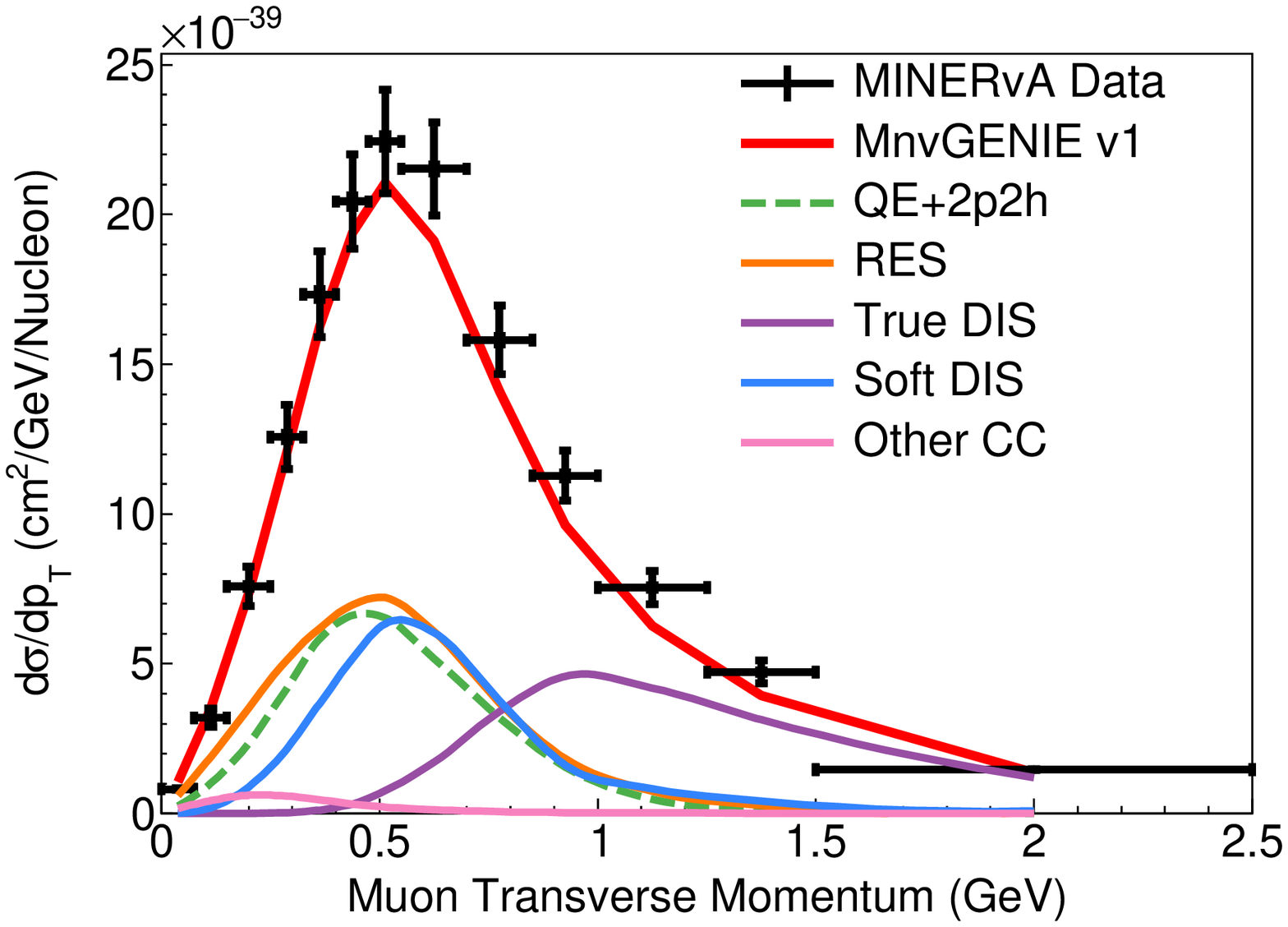} 
	\includegraphics[width=0.95\linewidth]{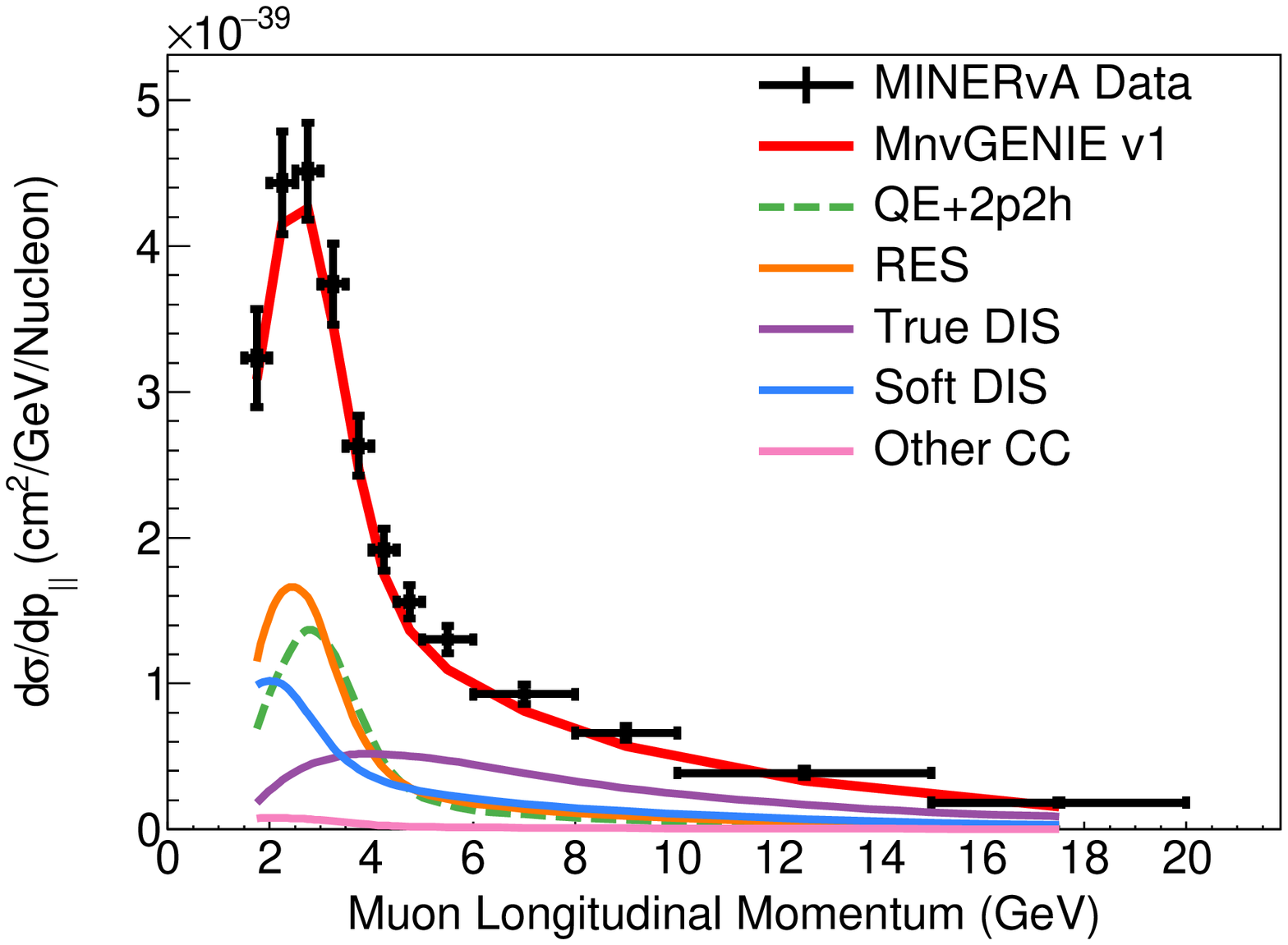} 
	\caption{Measured differential cross sections in transverse and longitudinal momenta. \tuneplot~is shown with its unstacked components.}
	\label{fig:ptpzsinglediff}
\end{figure}%
\begin{figure*}[p]
	\centering
	\includegraphics[width=0.95\linewidth]{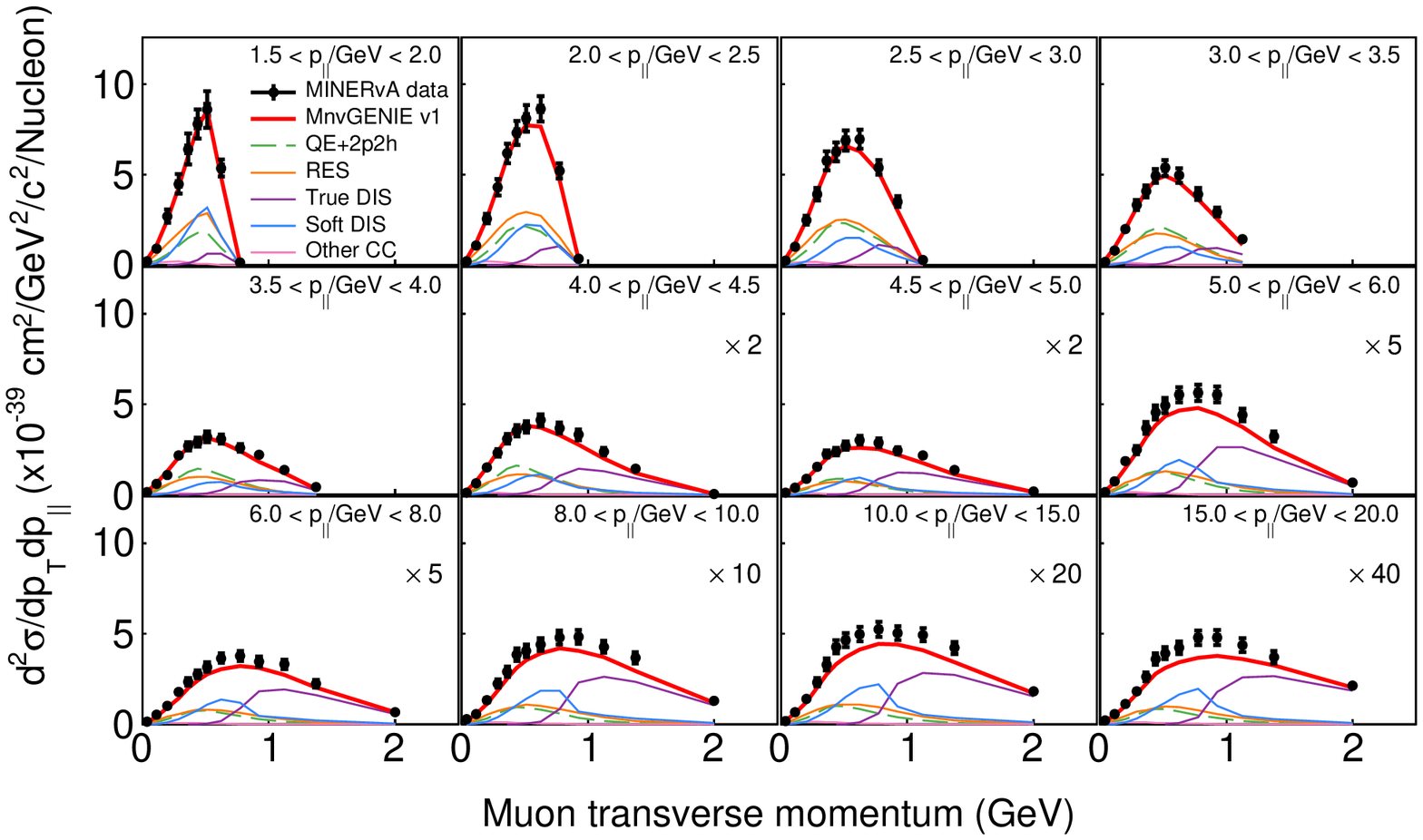} 
	\includegraphics[width=0.95\linewidth]{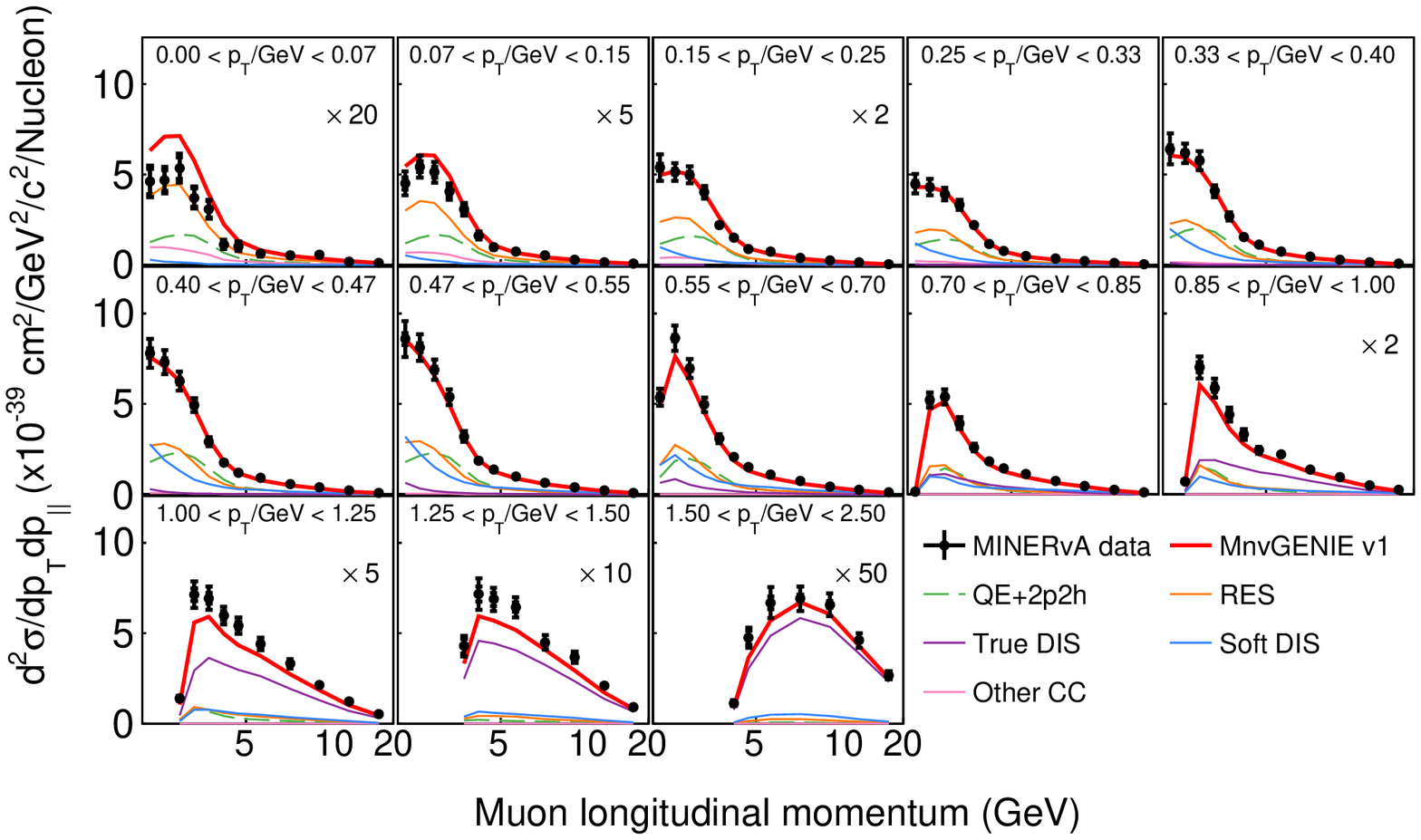} 
	\caption{Measured double-differential cross section in slices of \pz (top) and \pt (bottom). \tuneplot~is shown with its unstacked components.}
	\label{fig:ptpzdoublediff}
\end{figure*}
\begin{figure*}[p]
	\centering
	\includegraphics[width=0.95\linewidth]{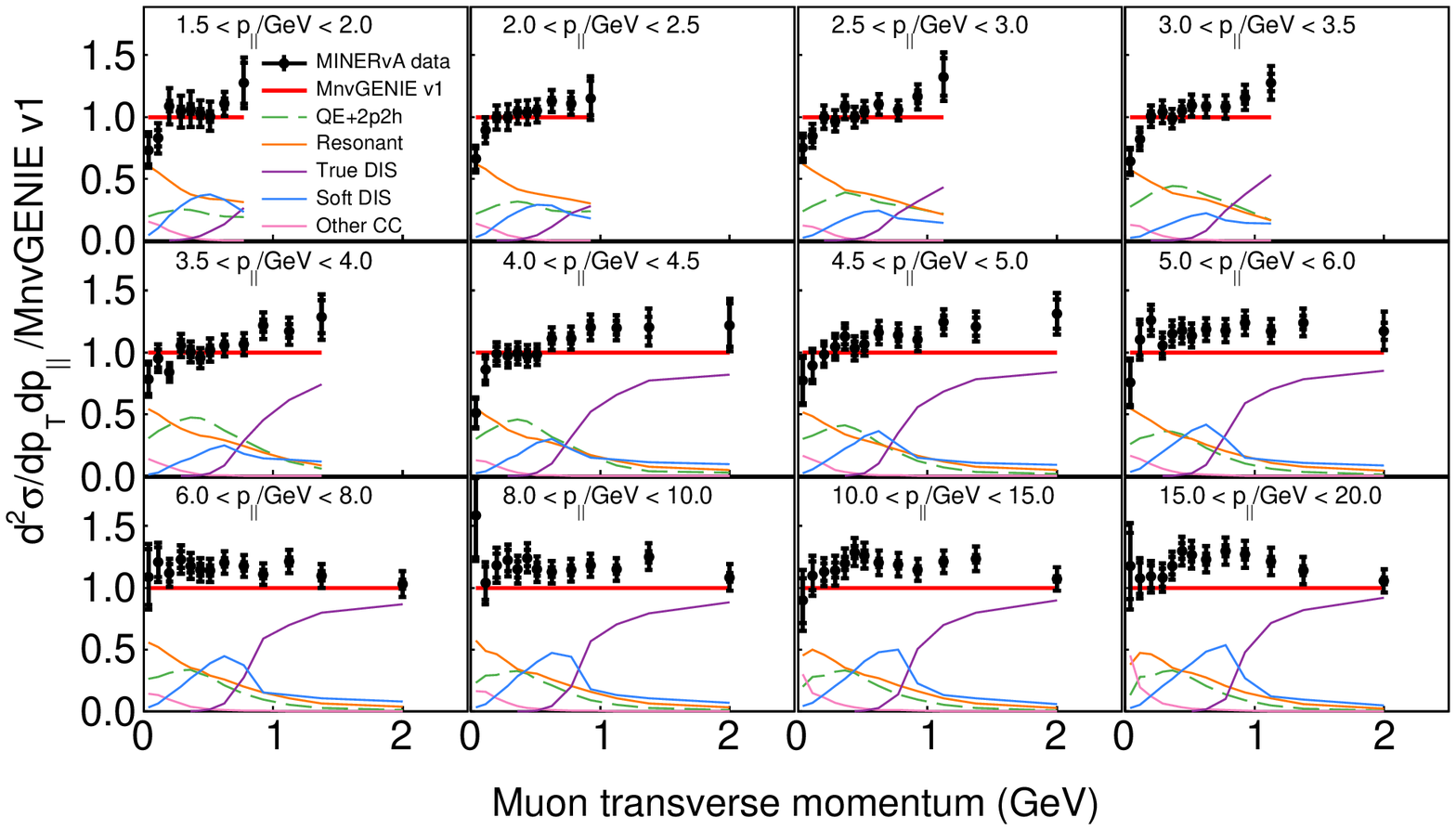} 
	\includegraphics[width=0.95\linewidth]{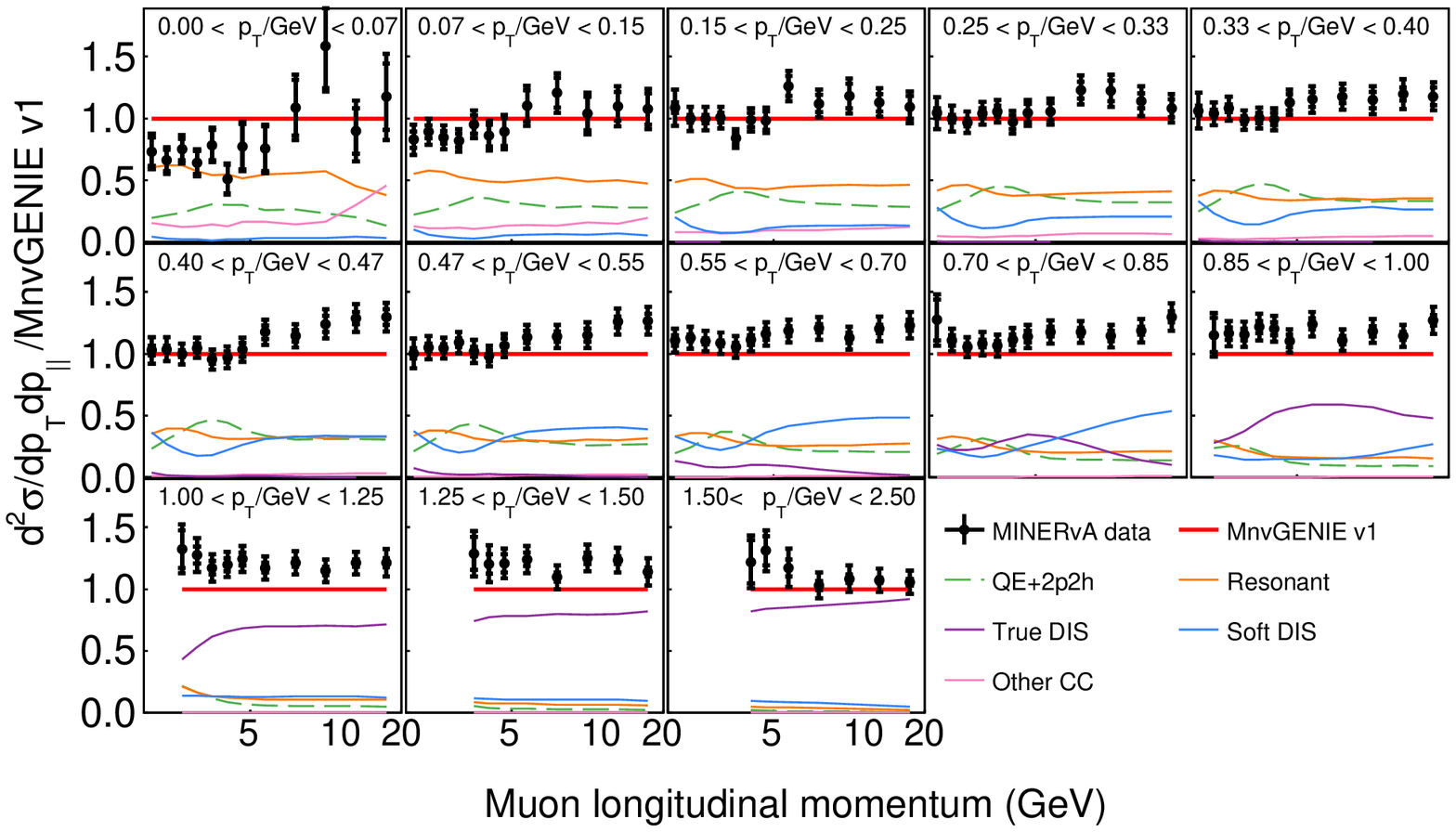} 
	\caption{Measured double-differential cross section as a ratio to \tuneplot~in slices of \pz (top) and \pt (bottom). Unstacked components of \tuneplot~are also shown as a ratio to the total MC.}
	\label{fig:ptpzdoublediffratio}
\end{figure*}

\section{Systematic Uncertainties}
\label{systematics}
A breakdown of systematic uncertainties on the single-differential cross sections is shown in Fig. \ref{fig:1dsys}. Both projections have similar uncertainties and contributions, with the flux being the dominant systematic uncertainty in both, and total uncertainties ranging from 8 to 10\%. 

An uncertainty summary for the double-differential cross sections is shown in Fig. \ref{fig:sys_ptpz}. Flux is again the dominant systematic, contributing at the 7\% level throughout the phase space. The uncertainty in the muon energy scale, the dominant component of the muon reconstruction uncertainty, is comparable to the flux uncertainty at low longitudinal momentum.
There is a noticeable effect in the lowest \pz~bin in which the muon energy uncertainty fluctuates between high and low points. This effect appears as a result of unfolding, with higher numbers of iterations producing highly anti-correlated bins. The other systematics categories each make up a smaller contribution to the total uncertainty than the statistical uncertainty.
The model uncertainties are evaluated through each stage of the cross section extraction using the GENIE reweighting framework~\cite{Andreopoulos:2009rq}.

\section{Results}
\label{sec:results}
\subsection{Interaction channel model components}
The extracted single-differential cross sections in longitudinal and transverse momentum are shown alongside \tune~in Fig.~\ref{fig:ptpzsinglediff}. This figure also shows the unstacked interaction channel components predicted by \tuneplot. The transverse momentum projection shows a separation of true DIS type events, but the QE+2p2h, RES, and soft DIS interaction channels are all occupying the same area of \pt. \tuneplot~
makes an underprediction from 0.55\,\textless\,\pt\,\textless\,\unit[1.5]{GeV}, with agreement within 1$\sigma$ in the mid-\pt~and highest-\pt~bin. 
The muon longitudinal momentum projection shows very little separation between any of the interaction channels. In this projection there is agreement with \tuneplot~in the first few bins, with an underprediction of the cross section for all longitudinal momenta greater than \unit[4]{GeV}.

\begin{figure}[tp]
	\centering
	\includegraphics[width=0.95\linewidth]{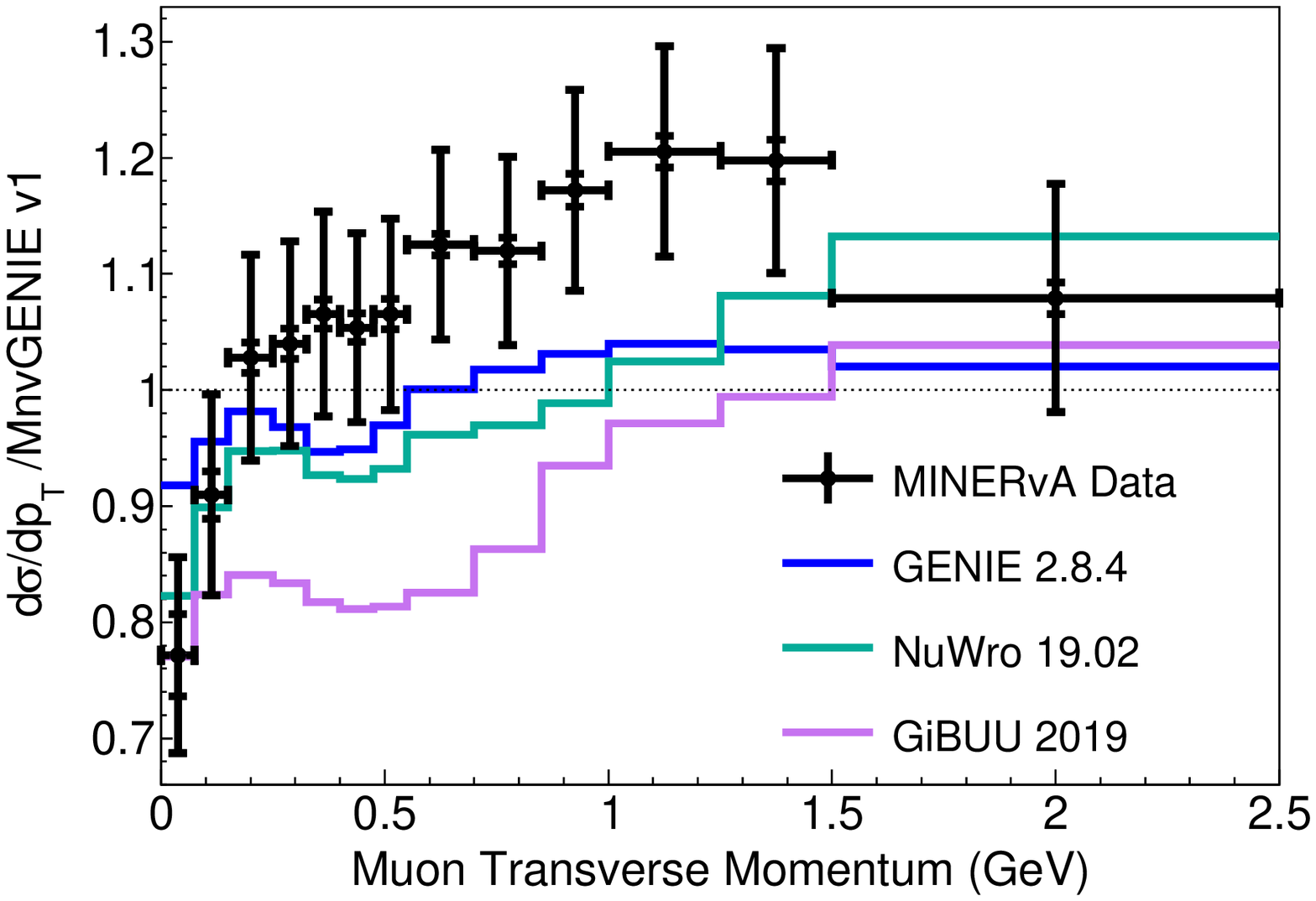} 
	\includegraphics[width=0.95\linewidth]{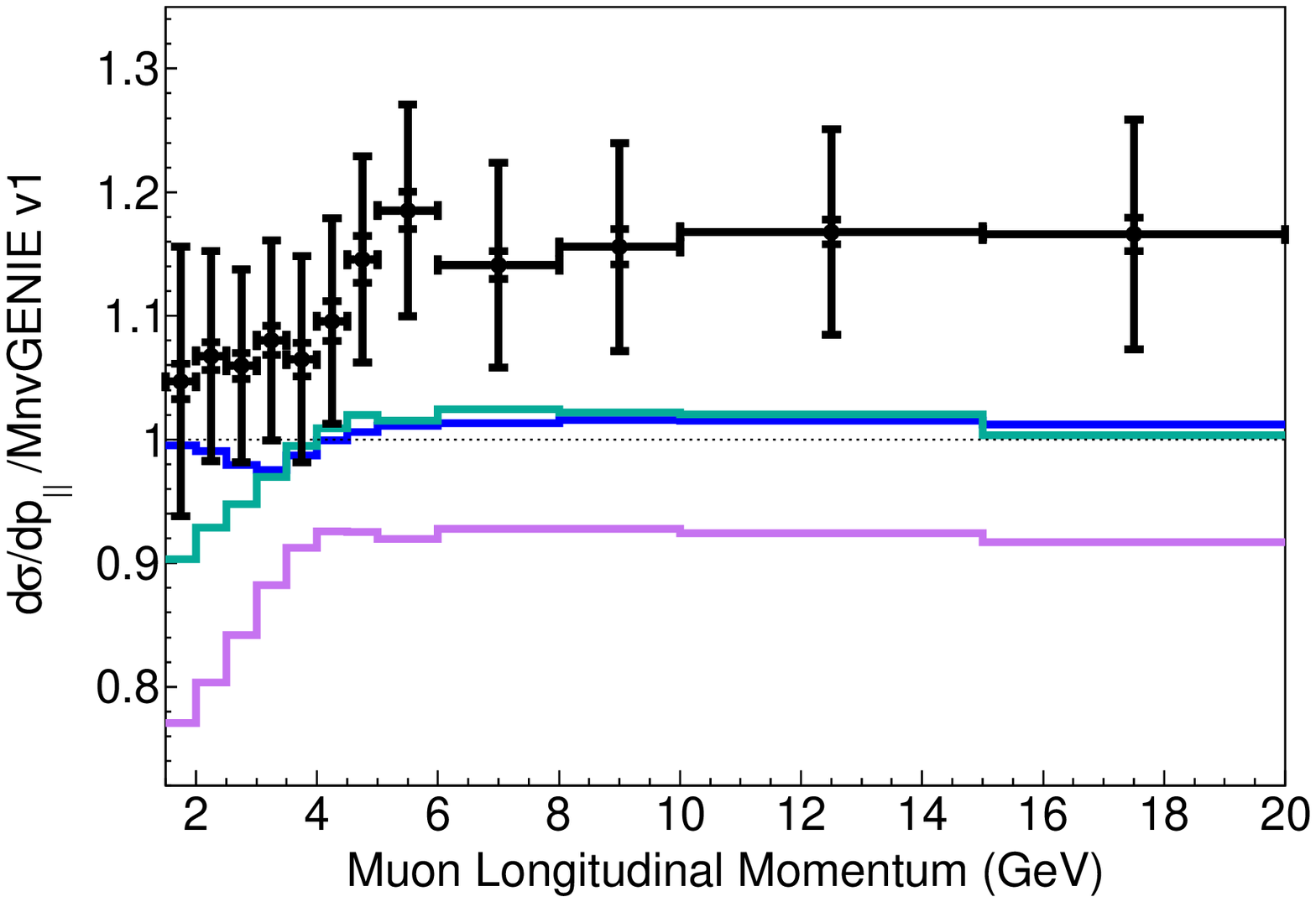} 
	\caption{Absolutely normalized ratios of data, GENIE 2.8.4, NuWro, and GiBUU to \tuneplot~for \pt~and \pz. The transverse momentum projection shows tension between all models and data in the 0.55\,\textless\,\pt\,\textless\,\unit[1.50]{GeV} range, with the highest \pt~bin modeled the best. In longitudinal momentum, all models underpredict the cross section, with the most significant discrepancy of a 20 to 40\%  normalization difference occurring with GiBUU.}
	\label{fig:singlediffratio}
\end{figure}%
\begin{figure}[tp]
	\centering
	\includegraphics[width=0.95\linewidth]{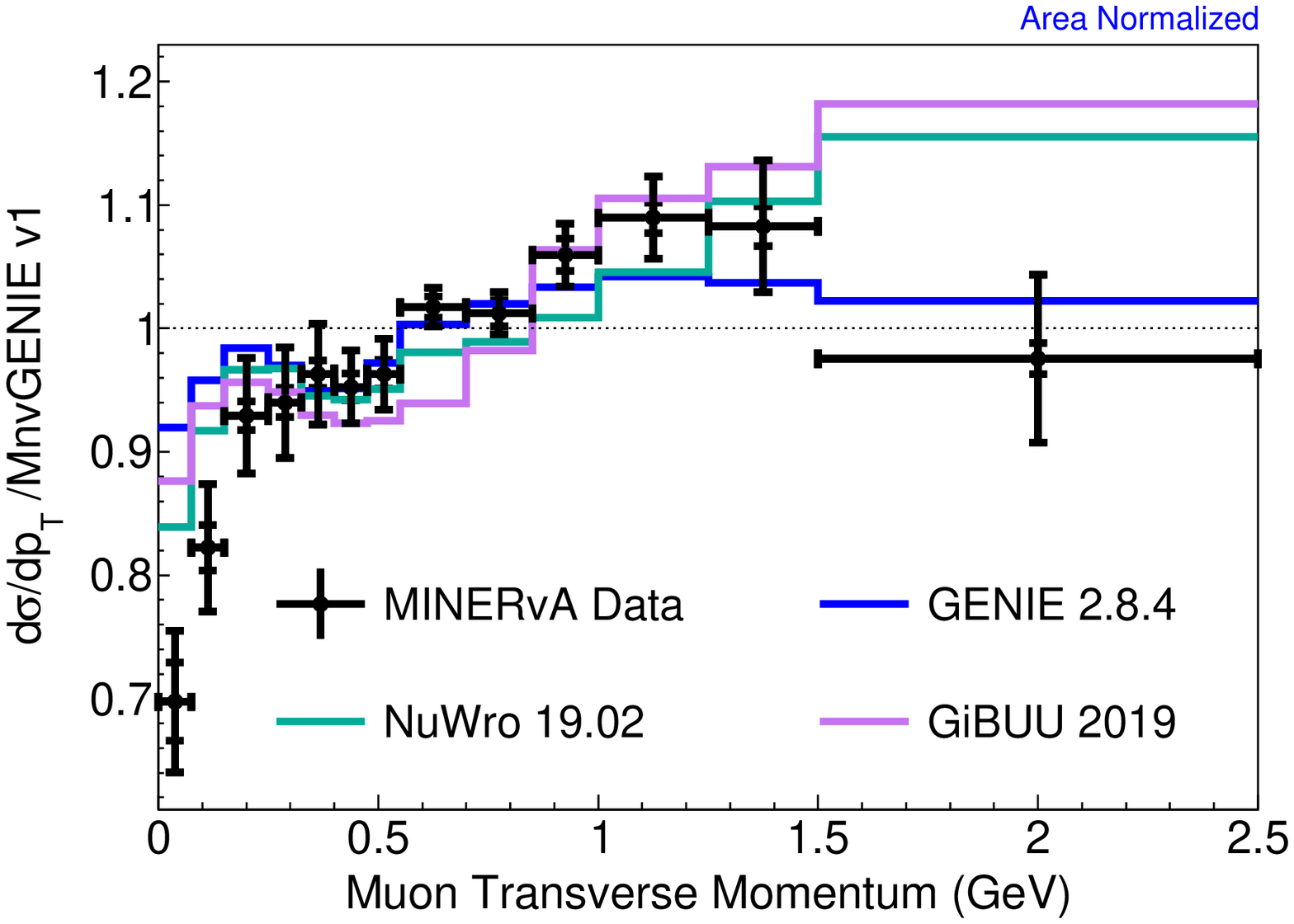} 
	\includegraphics[width=0.95\linewidth]{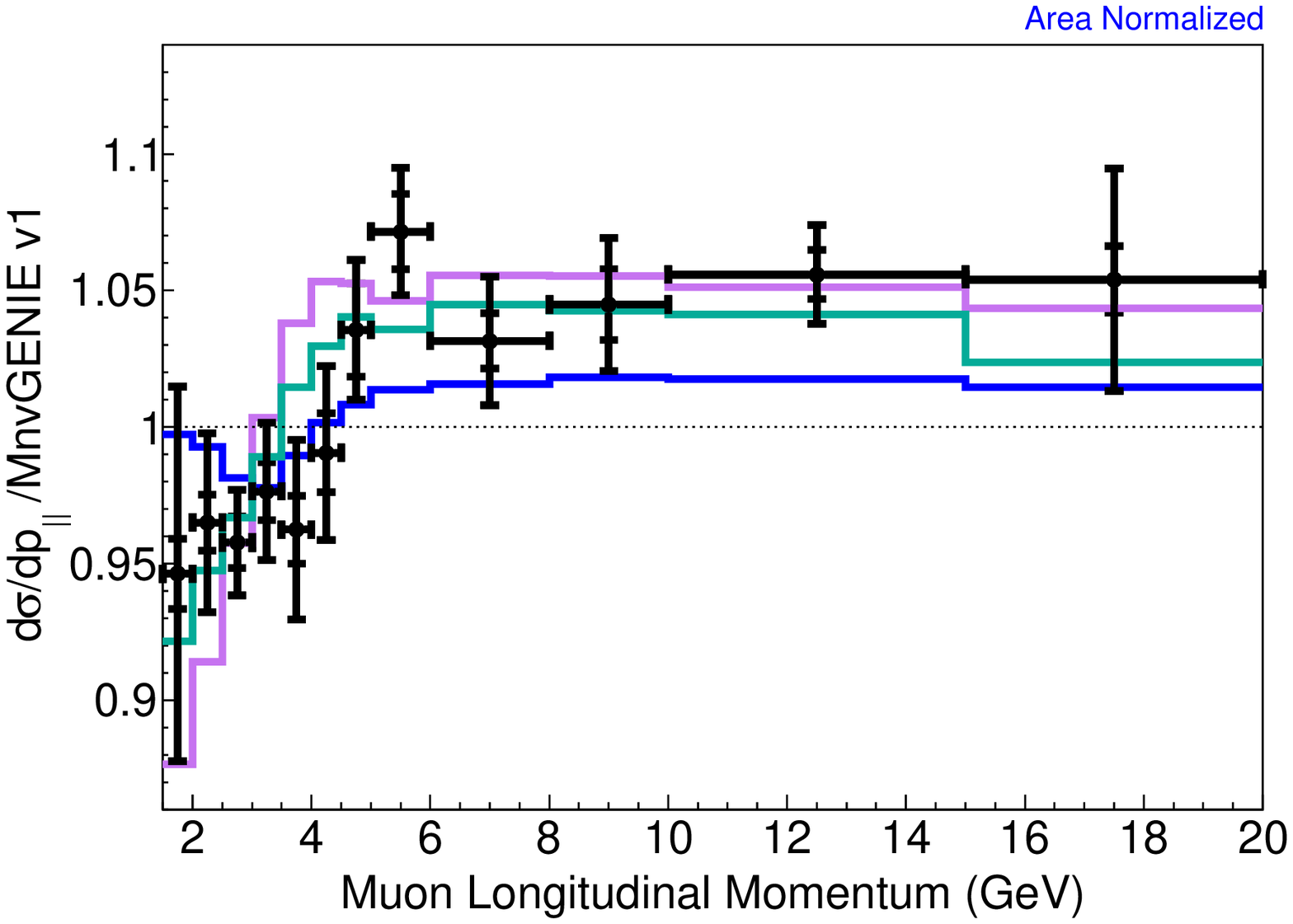} 
	\caption{Shape-only ratios of data, GENIE 2.8.4, NuWro, and GiBUU to \tuneplot~ for \pt~and \pz. In the transverse momentum projection GENIE 2.8.4 performs the best, agreeing well with data for all but the lowest \pt~bins. In longitudinal momentum, GENIE 2.8.4 and NuWro show the best agreement, with 75\% of bins in agreement.}
	\label{fig:singlediffratioarea}
\end{figure}

The double-differential cross section is shown in Fig.~\ref{fig:ptpzdoublediff} along with \tuneplot~and an unstacked breakdown of the simulated interaction types. This double-differential result shows much better separation between interaction channels than shown in either single-differential projection. 
Some notable features of this double-differential result include an overprediction of the cross section in the majority of the 0 to \unit[0.07]{GeV} \pt~ bins, an underprediction of the cross section for bins with high \pz~and midrange \pt, and underpredictions for \pt\textgreater \unit[0.85]{GeV} and \pz\,\textless\,\unit[6.0]{GeV}. The data-MC differences do not track with any individual interaction channel, as can be more clearly seen in Fig.~\ref{fig:ptpzdoublediffratio}, which shows the data and simulated interaction types as a ratio to \tuneplot. 
One channel which preforms comparatively poorly is the soft DIS; when it is the dominant interaction channel, \tuneplot~consistently underpredicts the cross section. Some portions of phase space in which true DIS is the dominant channel see a similar trend, specifically in the \pt~range from \unit[0.85]{GeV} to \unit[1.50]{GeV}. However, in bins with higher average values of W, specifically the bins with an average W\,\textgreater\,\unit[3.5]{GeV} (the 5 highest \pz~bins with \pt\,\textgreater\,\unit[1.50]{GeV} and the highest \pz~bin with \unit[1.25]{GeV}\,\textless\,\pt\,\textless\,\unit[1.50]{GeV}), all have true DIS contributions of greater than 80\% and show good agreement with the data.

\begin{figure*}[p]
	\centering
	\includegraphics[width=0.95\linewidth]{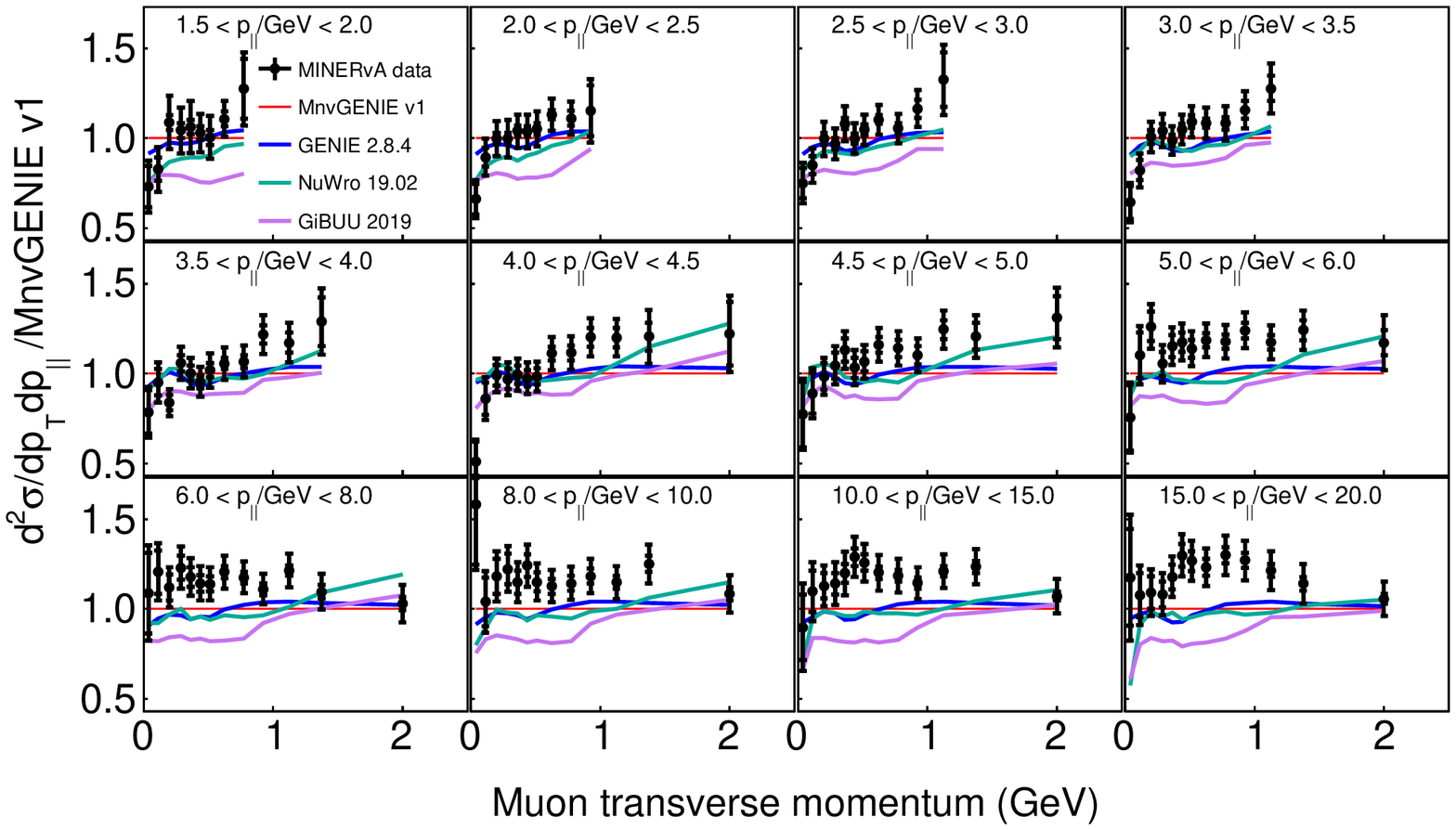}
	\includegraphics[width=0.95\linewidth]{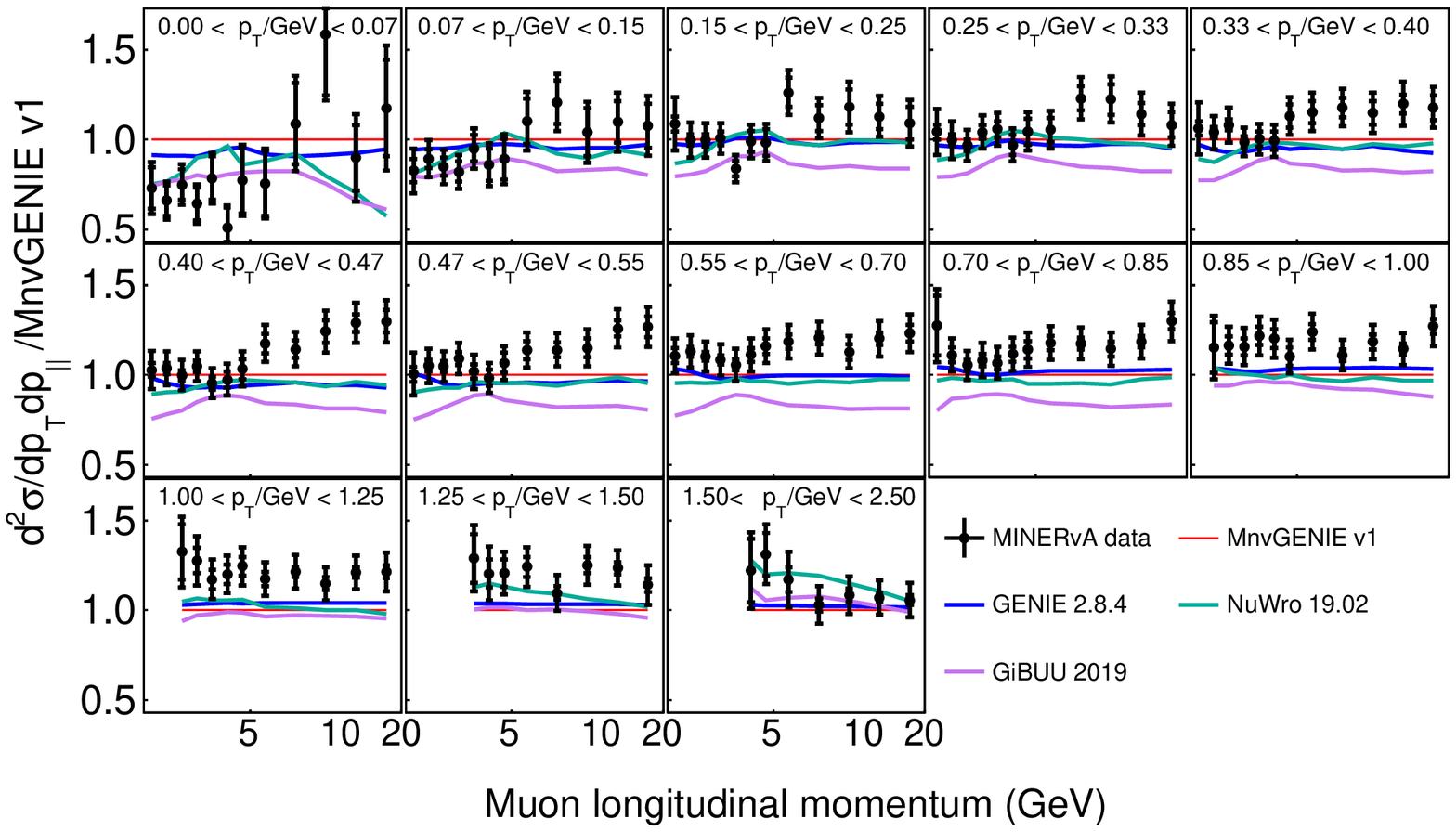}
	\caption{Ratios of the measured cross section, untuned GENIE 2.8.4, NuWro 19.02, and GiBUU 2019 to \tuneplot. None of these models are able to faithfully reproduce the measured cross sections throughout the two dimensional phase space. The region which has the best model agreement is in the lower half of the \pz~range with 0.15\,\textless\,\pt\,\textless\,\unit[0.55]{GeV}.}
	\label{fig:modcomp_modelsgroup}
\end{figure*}

\begin{figure*}[p]
	\centering
	\includegraphics[width=0.95\linewidth]{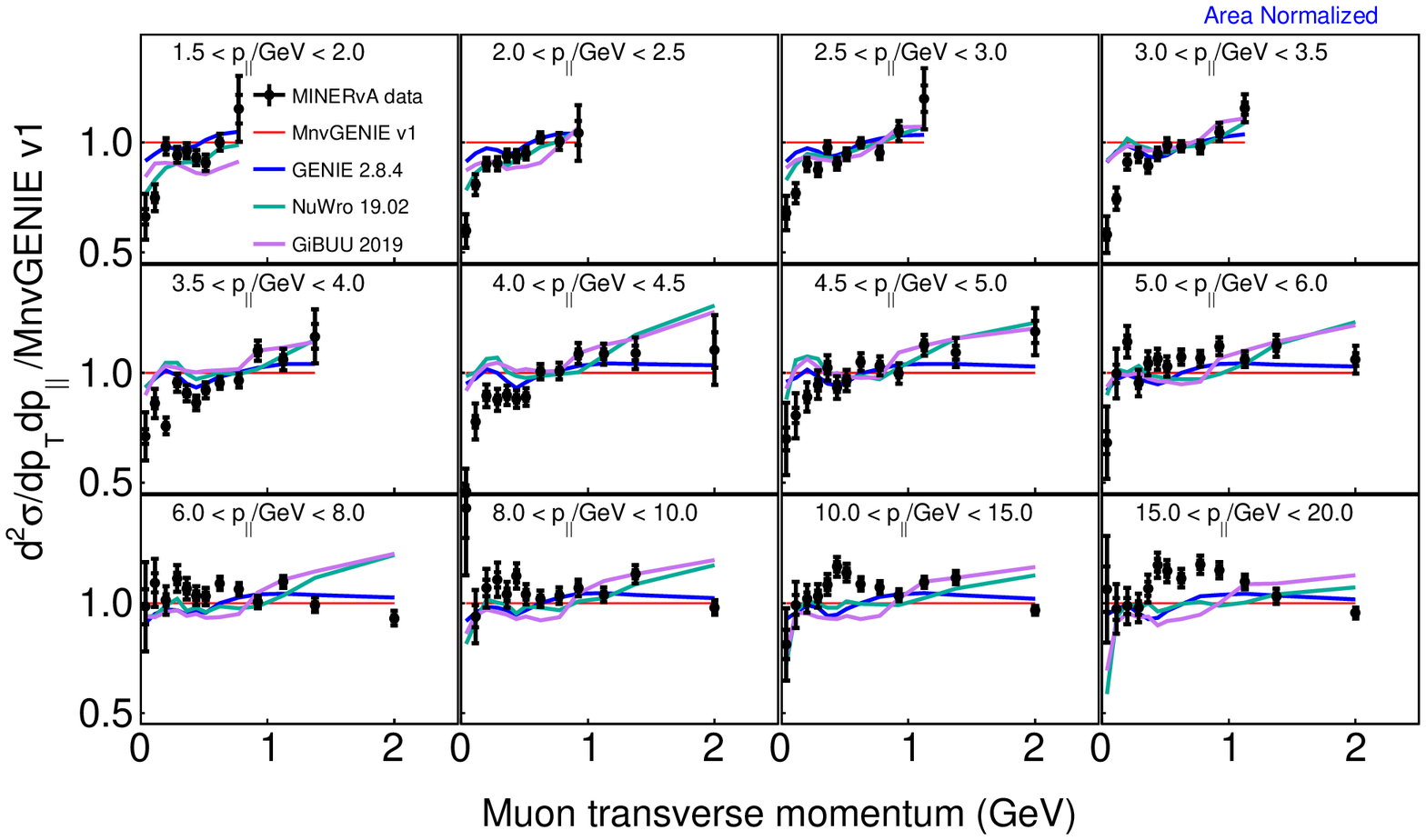}
	\includegraphics[width=0.95\linewidth]{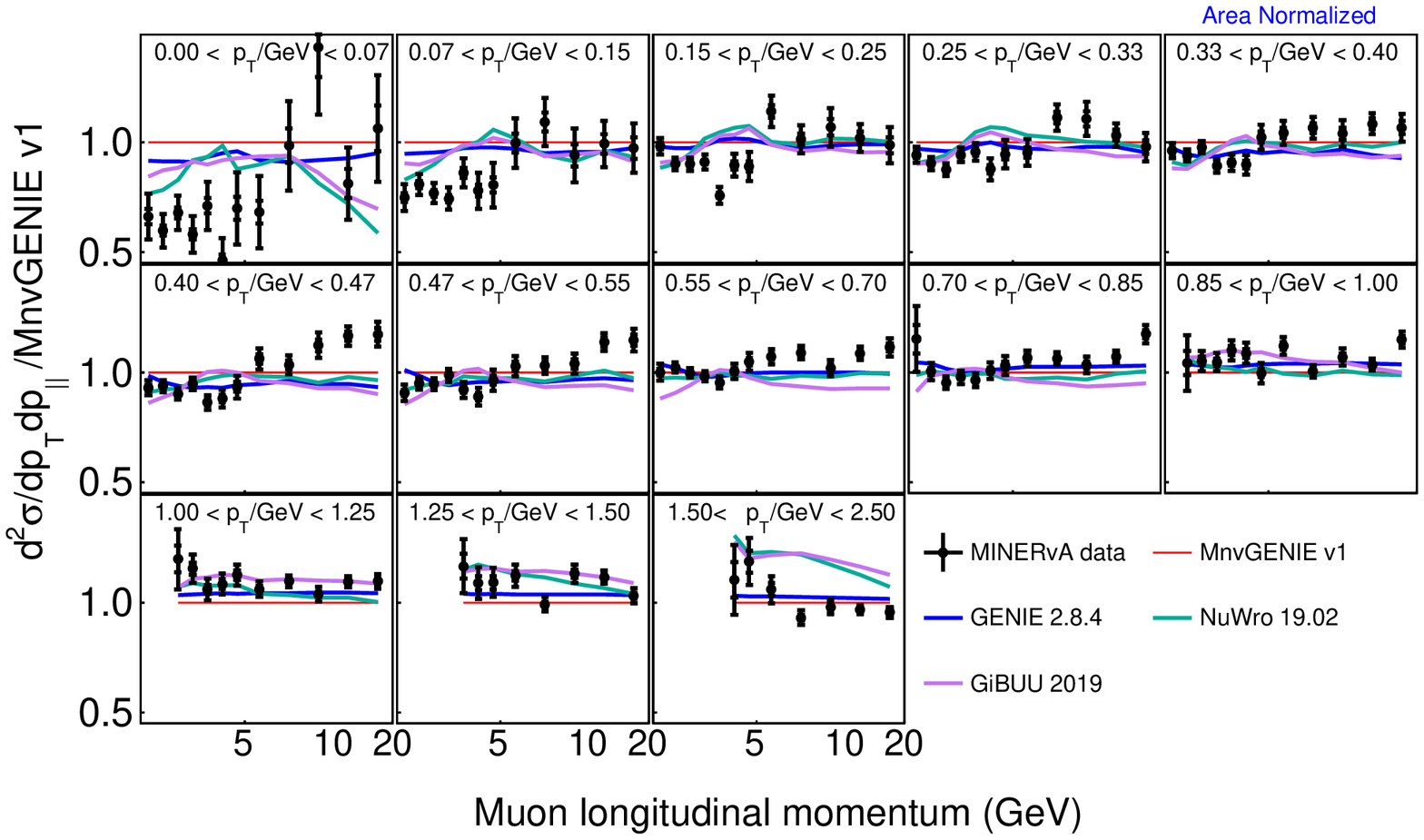}
	\caption{Shape-only ratios of the measured cross section, untuned GENIE 2.8.4, NuWro 19.02, and GiBUU 2019 to \tuneplot. Data in the region 2.0\,\textless\,\pz\,\textless\,\unit[5.0]{GeV} with \pt\,\textless\,\unit[0.25]{GeV} has notable tension with these models, with very few data bins exhibiting 1$\sigma$ agreement with any model.}
	\label{fig:modcomp_modelsgrouparea}
\end{figure*}

\begingroup
\squeezetable
\begin{table}[tp]
	\begin{tabular}{lcr}
		\hline
		Process Variant & Standard $\chi^{2}$ & Log-normal $\chi^{2}$ \\ 
		\hline \hline
		\tuneplot & 495 & 547 \\ \hline
		GENIE 2.8.4 & 422 & 491 \\ \hline
		\tunevT & 475 & 665 \\ \hline
		GENIE + piontune & 477 & 580 \\ \hline
		GENIE + RPA & 327 & 459 \\ \hline
		GENIE + RPA + 2p2h & 402 & 464\\ \hline
		GENIE + 2p2h & 690 & 725 \\ \hline
		\tune +  MINOS $\pi$ low \qsq~sup. & 381 & 526 \\ \hline
		\tuneplot + nCTEQ15 DIS & 503 & 551 \\ \hline
		\tuneplot + nCTEQ$\nu$ DIS & 506 & 565 \\ \hline
		\tuneplot + AMU DIS & 549 & 636 \\ \hline
		NuWro & 820 & 587 \\ \hline 
		GiBUU & 767 & 815 \\ \hline 
	\end{tabular}
	\caption{The $\chi^{2}$ calculated via standard and log normal calculations for each model and model variant used in this analysis. There are 144 degrees of freedom.}
	\label{tab:DDModelComp}
\end{table}
\endgroup

\subsection{Performance comparisons of neutrino event generators}
A summary of \xsq~values for each model and tune is shown in Table \ref{tab:DDModelComp}. This table presents the sum of bin-by-bin $\chi^2$ calculated for the double-differential result with a full treatment of correlations, using both standard and log normal calculations. GENIE with the addition of the nonresonant pion tune and quasielastic random phase approximation (GENIE+RPA), has the lowest $\chi^2$ values of 327 and 459, for standard and log normal calculations respectively, with 144 degrees of freedom. High values of \xsq/DoF have similarly been seen for prior measurements of double-differential cross sections, such as a \qelike~measurement by \minerva~\cite{Ruterbories:2018gub}, and inclusive measurements made by T2K ~\cite{Abe:2018uhf} and \uboone~\cite{Adams:2019iqc}.

GENIE 2.8.4, NuWro and GiBUU are compared to the data by taking ratios of both to \tuneplot, as single-differential projections in Fig.~\ref{fig:singlediffratio}.  In the longitudinal momentum projection, all of the neutrino generators used tend to underpredict the cross sections at high longitudinal momentum. All of the generators, with the exception of GiBUU, underpredict the data in this area by approximately 10 to 15\%. GiBUU shows the largest discrepancy, with a 10 to 20\% normalization difference with respect to the other models, resulting in a 20 to 40\% absolute normalization difference with the data. In the transverse momentum projection, the highest bin, ranging from 1.5\,\textless\,\pt\,\textless\,\unit[2.5]{GeV}, is the best-modeled, with all 4 models in agreement with the data. GENIE 2.8.4 and NuWro both agree with the majority of the data bins for \pt\,\textless\,\unit[0.33]{GeV}. \tuneplot~has the best agreement in the range of 0.15\,\textless\,\pt\,\textless\,\unit[0.55]{GeV}, and GiBUU has the worst agreement with only three bins being consistent with data.
\begin{figure}[tp]
	\centering
	\includegraphics[width=0.95\linewidth]{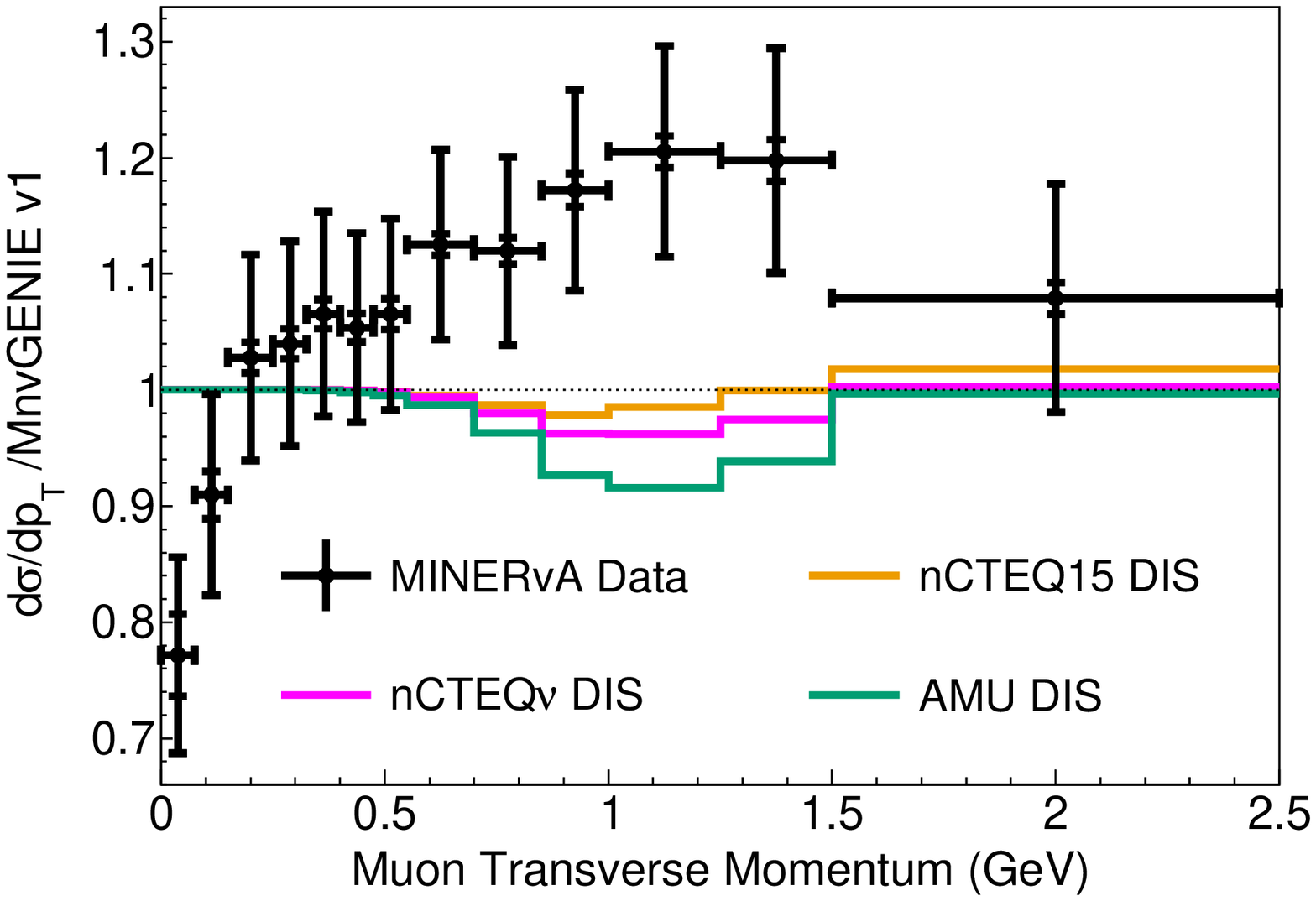} 
	\includegraphics[width=0.95\linewidth]{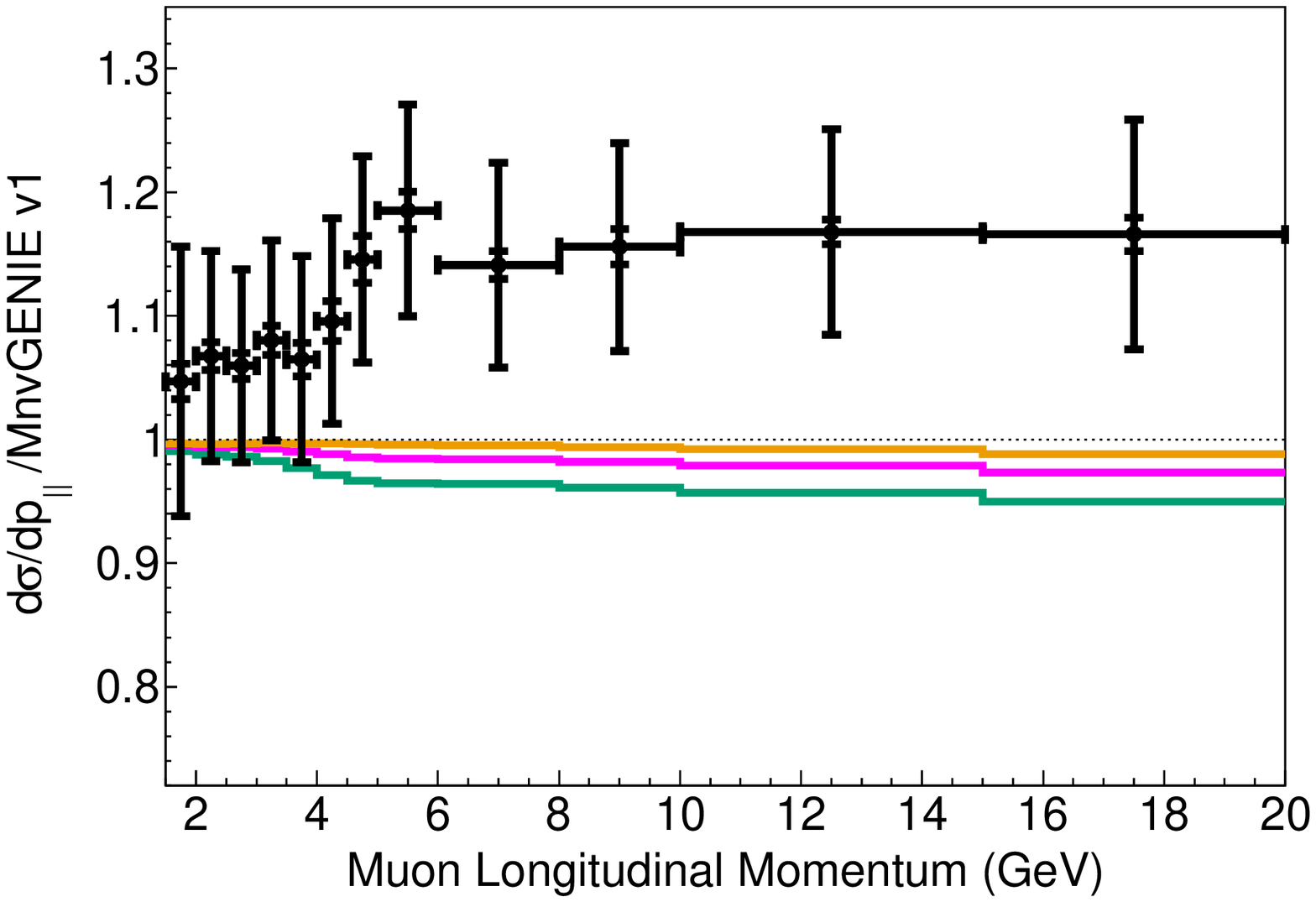} 
	\caption{Versions of \tuneplot~altered to use nCTEQ15, nCTEQ$\nu$, and AMU true DIS models are shown alongside data as ratios to \tuneplot~for \pt~and \pz. All of these models tend to underpredict the cross sections in  all areas of sizable DIS contributions except the highest \pt~bin.}
	\label{fig:singlediffdisratio}
\end{figure}

When making shape comparisons, however, as shown in Fig.~\ref{fig:singlediffratioarea}, GENIE 2.8.4 has the best agreement with the data in the transverse momentum projection, having a majority of bins 
within 1$\sigma$. GENIE 2.8.4, NuWro and GiBUU all match the shape of the data in the 0.25\textless\,\pt\,\textless\,\unit[0.55]{GeV} bins. \tune~performs the worst in the transverse momentum shape comparison, with only three bins in agreement with the data. 
For the shape-only longitudinal momentum model comparisons, NuWro and GENIE 2.8.4 have the best agreement, each with only a few scattered bins not in agreement. 
In neither projection does \tuneplot~stand out as being a particularly good fit to the data. 

The full double-differential cross-section ratios for these three event generators are shown in Fig.~\ref{fig:modcomp_modelsgroup}. Again, none of these models have good agreement with the data throughout the full phase space. The midrange \pt~shows the same GiBUU normalization difference seen in the longitudinal momentum projection. NuWro has the best agreement at high \pt, with all but one of the highest \pt~bins in agreement, and the most bins in agreement in the second highest \pt~bin. The bins with \pz\,\textless\,\unit[5.0]{GeV} 
and 0.15\,\textless\,\pt\,\textless\,\unit[0.55]{GeV} are among the best modeled, with \tuneplot~in agreement with data within 1$\sigma$ for 33 of these 35 bins and GENIE 2.8.4 with 83\% of bins in agreement. NuWro also has fairly good agreement in this region, especially for the subrange of 3.5\,\textless\,\pz\,\textless\,\unit[5.0]{GeV}, where 80\% of the bins are in agreement.
All of the models consistently underpredict the data in the region with a longitudinal momentum greater than \unit[5]{GeV} and 0.33\,\textless\,\pt\,\textless\,\unit[1.25]{GeV}, similarly seen in Fig.\ref{fig:ptpzdoublediffratio}.

An area normalized version of Fig.~\ref{fig:modcomp_modelsgroup} is shown in Fig.~\ref{fig:modcomp_modelsgrouparea}. The area normalization is applied as a single factor to all panels simultaneously for all of the double-differential results. The 7\% flux uncertainty is largely uniform, so the \xsq~calculated using the covaraince matrix partially accounts for such overall normalization effects. The area normalized \tuneplot, NuWro and GiBUU curves are scaled by normalization factors of 1.11, 1.13, and 1.26 respectively. The shape agreement is also poor for these models. NuWro and GiBUU model the shape at high \pt~with \pz\,\textless\,\unit[5.0]{GeV} better than \tuneplot, with  81\% and 94\% of the four highest \pt~bins in this range in agreement with data, respectively.

\subsection{Examination of DIS models}

\begin{figure}[tp]
	\centering
	\includegraphics[width=0.95\linewidth]{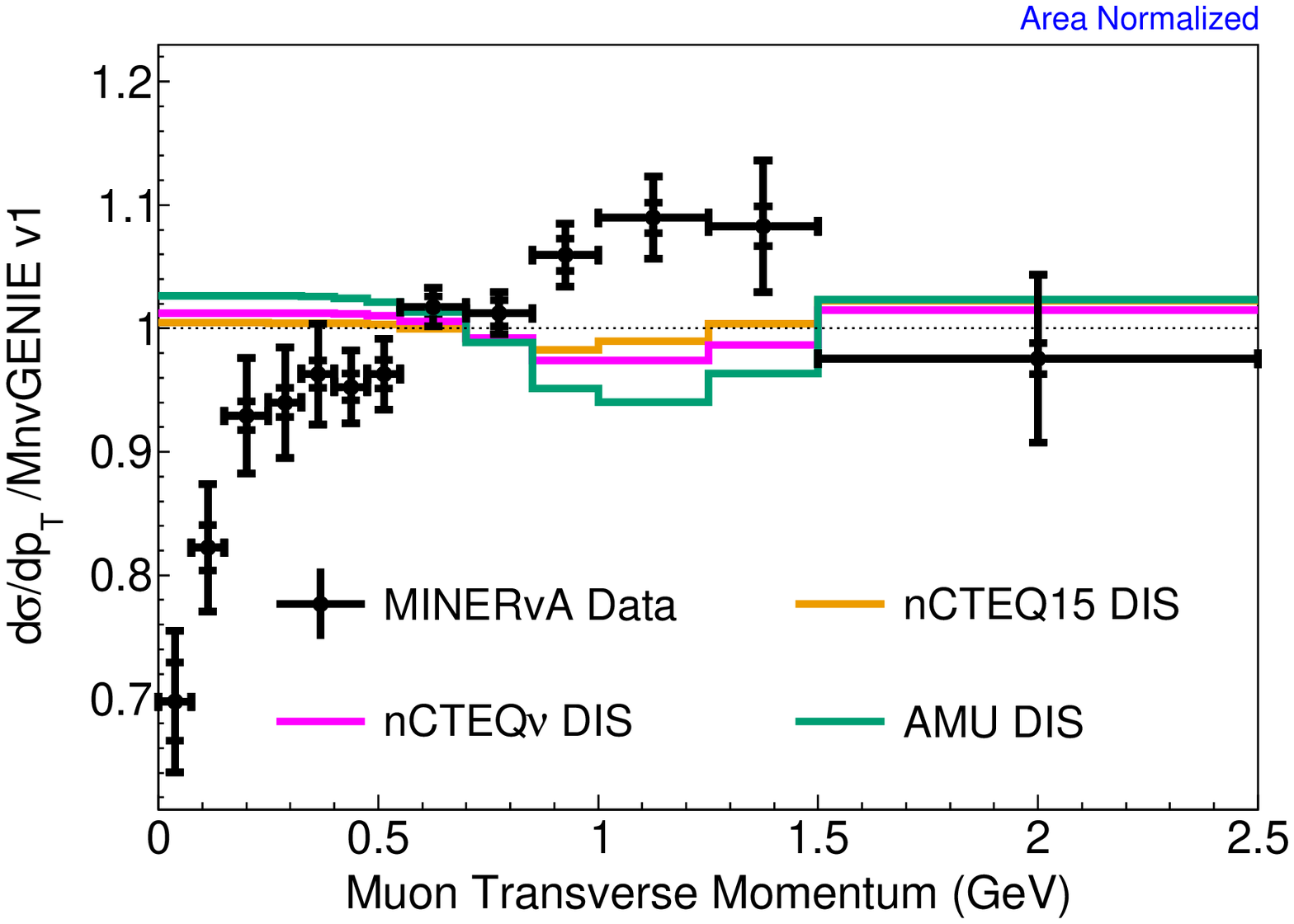} 
	\includegraphics[width=0.95\linewidth]{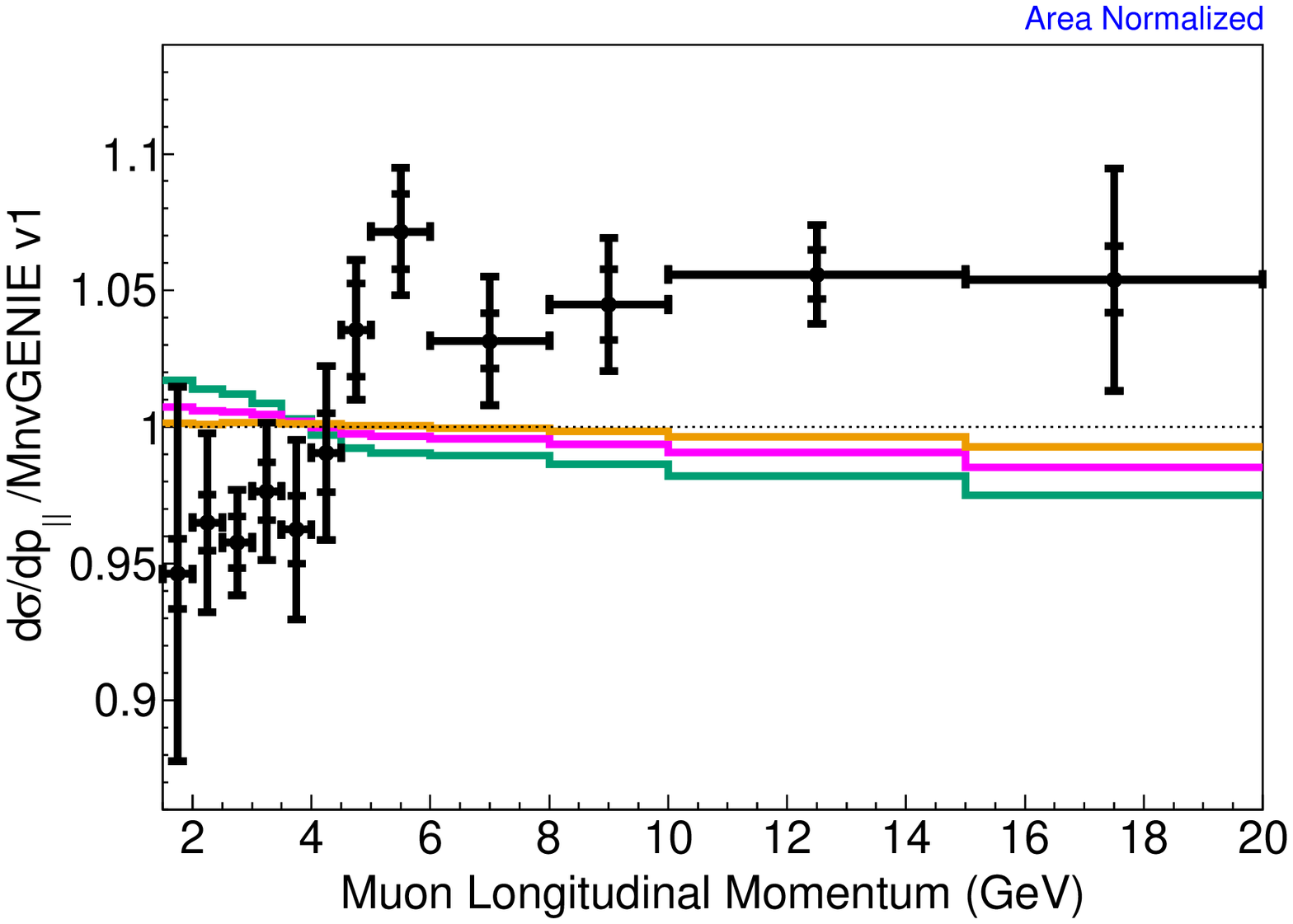} 
	\caption{Shape-only versions of \tuneplot~modified to use nCTEQ15, nCTEQ$\nu$ and AMU true DIS models as ratios to \tuneplot~for \pt~and \pz. In the mid \pt~range in which these models differ most from \tuneplot, the modifications made cause an increased shape discrepancy with the data.}
	\label{fig:singlediffdisratioarea}
\end{figure}
Single-differential DIS model comparisons to nCTEQ15, nCTEQ$\nu$ and AMU are shown in Fig.~\ref{fig:singlediffdisratio}. These comparisons use \tuneplot~with weights derived from the DIS models applied to only the true DIS ($W>\unit[2.0]{GeV}$, $Q^2>\unit[1.0]{GeV}$) component, as explained in Sec.~\ref{subsec:generators}. All of the resulting curves tend to underpredict the cross section in the areas with significant DIS contributions, except in the highest bin of transverse momentum. A shape-only version of this DIS model comparison is shown in Fig.~\ref{fig:singlediffdisratioarea}. All of these DIS models, when added to \tuneplot, show poor shape agreement with the data. 

\begin{figure*}[p]
	\centering
	\includegraphics[width=0.95\linewidth]{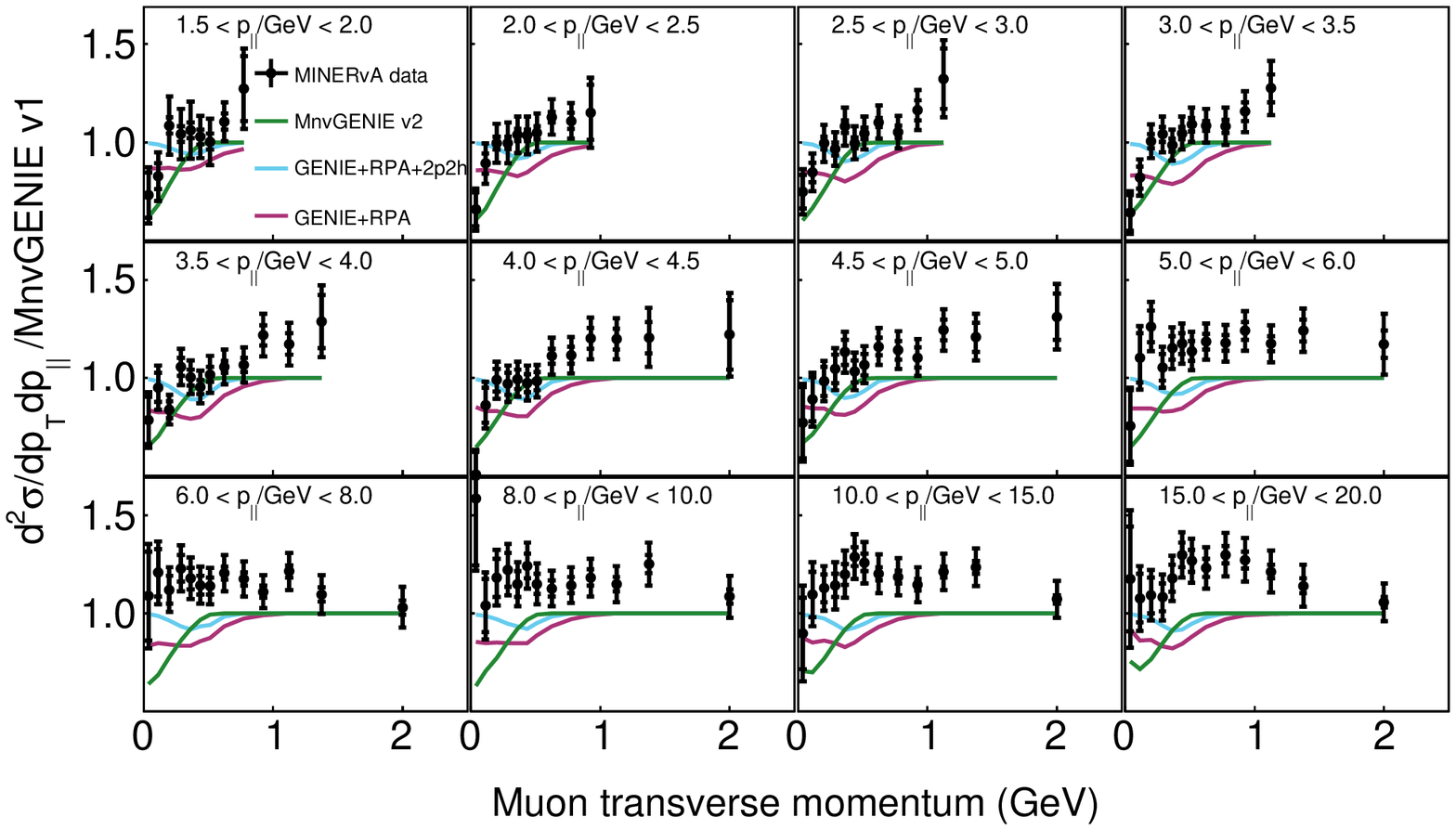}
	\includegraphics[width=0.95\linewidth]{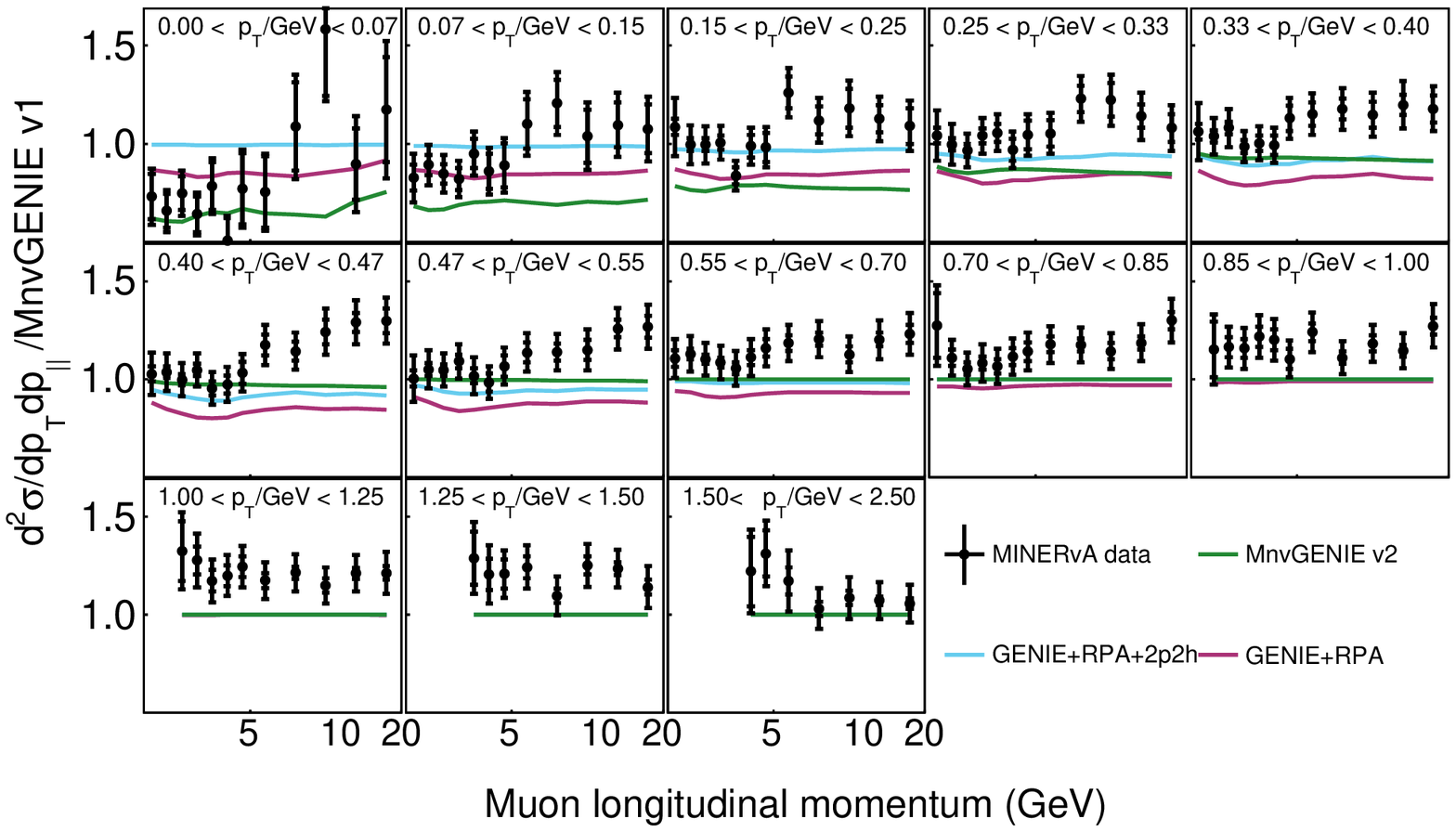}
	\caption{The ratio of data, \tunevT, GENIE with the additions of untuned 2p2h and RPA suppression, and GENIE with RPA suppression added, to \tuneplot. These three model variations provide some of the best \xsq~fits to the data.}
	\label{fig:modcomp_tunes}
\end{figure*}
\begin{figure*}[p]
	\centering
	\includegraphics[width=0.95\linewidth]{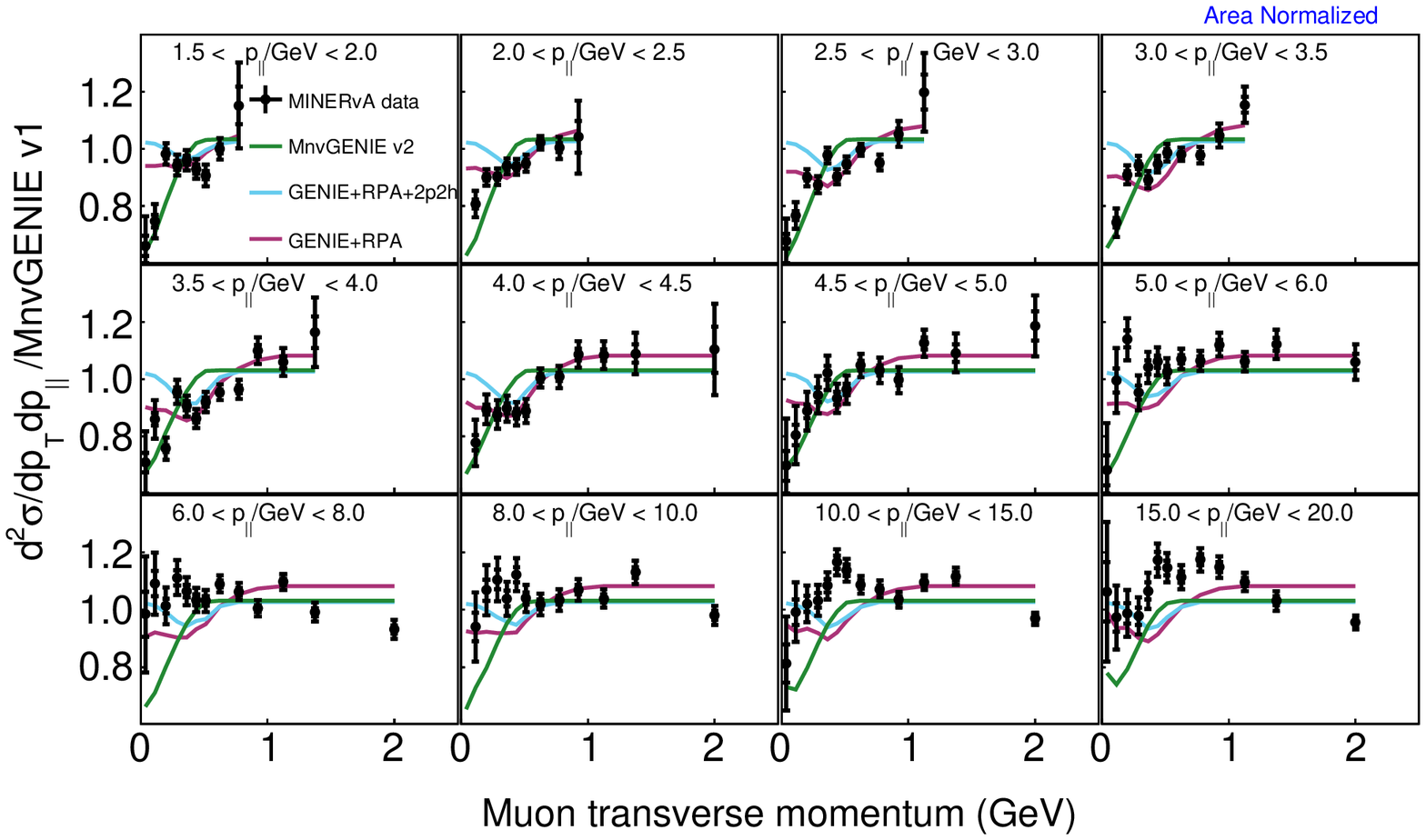}
	\includegraphics[width=0.95\linewidth]{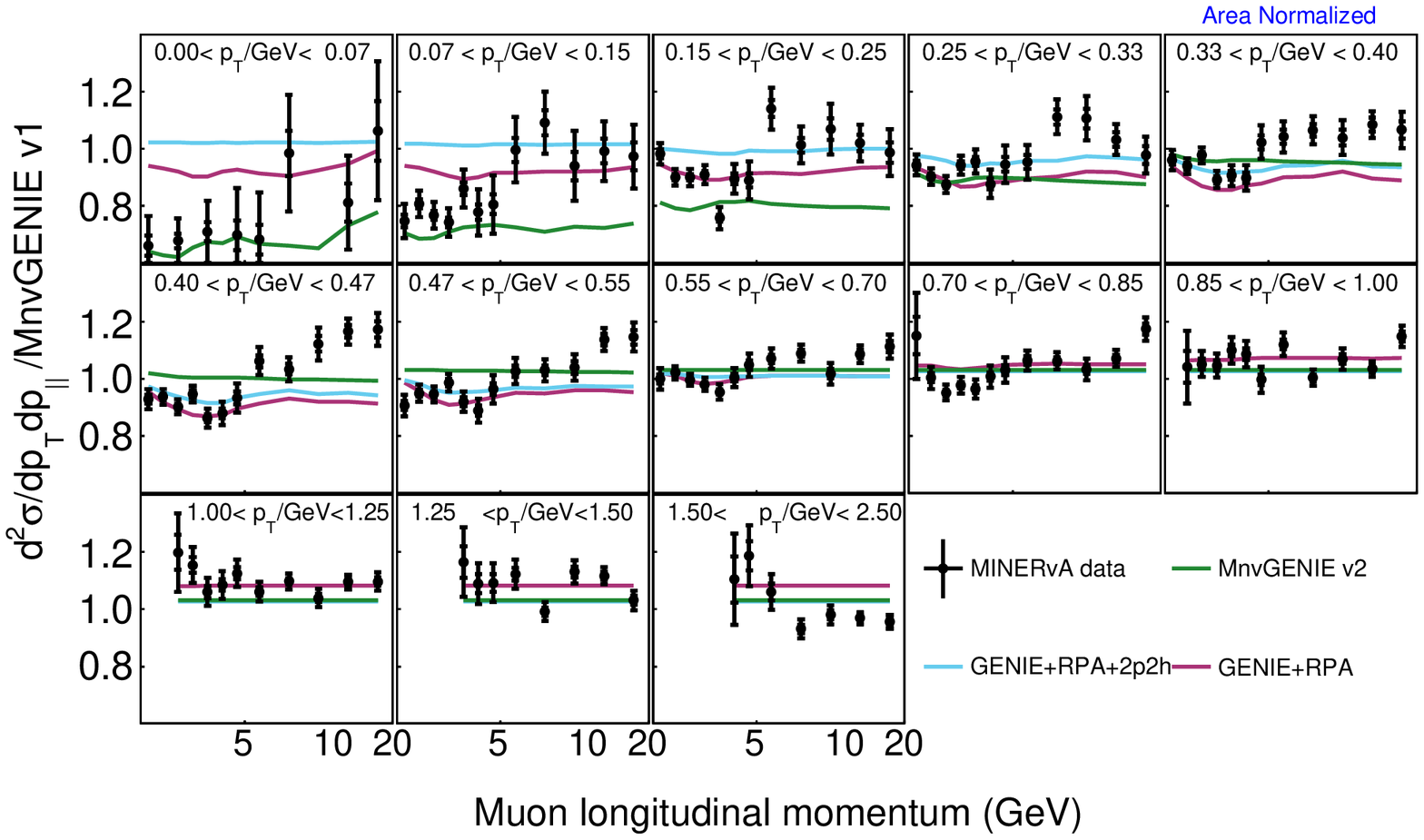}
	\caption{Shape-only ratio of data, \minerva~GENIE v2, GENIE with the additions of untuned 2p2h and RPA suppression, and GENIE with RPA suppression (and no 2p2h), to \tuneplot. The model variant with the best \xsq~fit, GENIE+RPA, reproduces the shape of the data in specific areas, for example in the first, second and sixth \pz~bins nearly all of the bins with \pt\,\textgreater\,\unit[0.15]{GeV}  are in agreement with data, but this variant also has many portions of phase space in which it fails to reproduce the shape of the data.}
	\label{fig:modcomp_tunes_area}
\end{figure*}

\subsection{Comparisons of Modeling Options with GENIE}
Various GENIE model variants are included in Table~\ref{tab:DDModelComp}. 
The addition of RPA, 2p2h and its tune to \minerva~data, and the suppression of low $Q^2$ resonances, are supported by comparisons using the measured hadronic system in \minerva.
Fig.~\ref{fig:modcomp_tunes} shows three model variants that have some of the lower \xsq~values. 
The first of these models is \tunevT, which includes addition of MINERvA's low momentum transfer resonance suppression described in Sec. \ref{subsec:generators}. This tune is identical to \tuneplot~at higher transverse momenta, with all of the differences occurring with \pt\,\textless\,\unit[1]{GeV}. The addition of low momentum transfer resonance suppression to MnvGENIE v2 does a reasonable job of reproducing the data in the first half of the \pz~bins, but maintains the large underprediction at higher longitudinal momentum, starting at approximately \unit[5.0]{GeV}. 
The shape of the suppression differs from the data trends; its addition generates decent agreement in the first transverse momentum bin, but is too strong in the second through fourth bins of transverse momentum.
When the MINOS version of this suppression is added to MnvGENIE v1, it produces better \xsq~fits than MnvGENIE v2. The MINOS suppression is similar to the MINERvA version in the lowest two \pt~bins, with a weaker suppression in higher \pt~bins. The latter difference produces better data agreement in those regions. 

The second-best log-normal \xsq~fit (third-best standard \xsq) is GENIE with the addition of quasielastic RPA suppression and Valencia model 2p2h, GENIE+RPA+2p2h.  This differs from \tuneplot~only in the 2p2h component, which is enhanced in \tuneplot, but not in GENIE+RPA+2p2h. 
For this reason, the region of interest for comparing these tunes is within the transverse momentum range of \unit[0.15]{GeV} to \unit[0.70]{GeV}, where all differences of significance occur. There is a slight dip in the data from 2.5\,\textless\,\pz\,\textless\,\unit[5]{GeV} for 0.25\,\textless\,\pt \,\textless\,\unit[0.40]{GeV}, which appears to slightly prefer the untuned 2p2h to the enhanced 2p2h used in \tuneplot. This effect is slightly more emphasized in the shape-only model comparisons in Fig.~\ref{fig:modcomp_tunes_area}.

Surprisingly, GENIE+RPA, which contains no 2p2h, 
is the model with the best \xsq. It shows a larger dip in the same area as GENIE+RPA+2p2h does, with a much larger effect at low longitudinal momentum, and extending further into low transverse momentum as well. In the absolutely normalized versions of these plots, the removal of 2p2h causes the model to dip substantially below the data in most areas of phase space (especially at higher longitudinal momenta). The shape agreement improves drastically with the removal of 2p2h in some regions; GENIE+RPA has data agreement in the majority of bins in the range from \unit[2.0]{GeV}\,\textless\,\pz \,\textless\,\unit[4.5]{GeV} with \pt\,\textgreater\,\unit[0.15]{GeV}, while \tuneplot~has poorer agreement in area normalized plots. As a best fit though, this model still fails to accurately produce the cross section shapes seen in the data across the full range of \pt~and \pz~and does a worse job at predicting the overall normalization than other models and tunes. 

Compared to \tuneplot, the description of the lowest \pt~bins improves with the removal of some event rate from at least one process. \tunevT~removes low $Q^2$ resonances, while GENIE+RPA instead removes all the 2p2h component. These defining characteristics of the two \minerva~tunes operate in overlapping regions of muon kinematics. The data may prefer future models with a modification of resonances more sophisticated then just a low $Q^2$ suppression.

The modifications of QE RPA suppression, 2p2h, enhanced 2p2h, and suppressing the low \qsq~resonance pion production are primarily motivated by \minerva~ data for the observed hadronic systems. This includes direct calorimetric measurements in \cite{Rodrigues:2015hik, Gran:2018fxa}, and the separation of samples with only protons and neutrons \cite{Ruterbories:2018gub, Patrick:2018gvi, Lu:2018stk, Cai:2019hpx} and with at least one pion \cite{Stowell:2019zsh, Le:2019jfy, Altinok:2017xua, McGivern:2016bwh, Eberly:2014mra}. Using the hadronic information in these ways provides relatively good separation of the QE, 2p2h, Delta resonance, and higher-W processes. The result is still an imperfect description of the muon kinematics in this new inclusive cross section, suggesting future focus on the detailed correlations between lepton kinematics and hadronic system.

\section{Conclusions}
This paper presents inclusive charged-current double- and single-differential cross sections in terms of longitudinal and transverse muon momentum. Measured cross sections are shown with comparisons to multiple variations of GENIE, in addition to model comparisons with NuWro and GiBUU, and DIS models nCTEQ15, nCTEQ$\nu$ and AMU.  The models see various levels of tension with the data, with no single model able to consistently reproduce the data throughout the two dimensional phase space. This poor agreement is seen both in absolutely normalized and shape-only model comparisons. 
There are some new models on the market which may be able to alleviate some of the tensions we see. These include other resonance models such as the MK model~\cite{Kabirnezhad:2017jmf}, alternative 2p2h models such as that from the SuSA group~\cite{Dolan:2019bxf}, other pion production models like those from Lyon~\cite{Martini:2014dqa}, Valencia~\cite{Hernandez:2013jka} and Ghent \cite{Gonzalez-Jimenez:2019qhq} groups, and different low \qsq~suppressions.

Models such as \tune~and MnvGENIE v2 were optimized to agree with previous \minerva~measurements in exclusive channels and limited kinematic regions~\cite{Rodrigues:2015hik}\cite{Ruterbories:2018gub}. They have been shown to see good agreement across different exclusive interaction channels~\cite{Patrick:2018gvi} and low-recoil samples~\cite{Gran:2018fxa}. However, the results presented here show that when all of these modifications are applied inclusively, having to contend with a large phase space with many contributing interaction channels, their predictive power is substantially diminished. 

Similarly, the suite of true DIS models used as partial model comparisons in this analysis were developed as theoretical and data-driven alternatives to other true DIS models such as those implemented in GENIE. However, these true DIS models do not result in better agreement than the GENIE DIS model. In fact, the addition of these true DIS models results in larger discrepancies with the data.

This measurement indicates that some form of a low $Q^2$ RES suppression helps to achieve better agreement in low \pt~regions, particularly for \pz\,\textless\,\unit[5.0]{GeV}. It also suggests that an enhancement of GENIE DIS may be called for in lower-W regions, because bins with an average W\,\textless\,\unit[3.5]{GeV} in which GENIE DIS is the dominant interaction channel show consistent underpredictions of the cross section.

The double- and single-differential cross sections show similar tensions with the model predictions.
These results demonstrate that improvements will need to be made to neutrino-interaction models if precision neutrino oscillation experiments hope to better constrain the systematics originating from cross section models. The measurements reported here should be helpful in validating these improvements.

\begin{acknowledgments}
	This document was prepared by members of the MINERvA Collaboration using the resources of the Fermi National Accelerator Laboratory (Fermilab), a U.S. Department of Energy, Office of Science, HEP User Facility. Fermilab is managed by Fermi Research Alliance, LLC (FRA), acting under Contract No. DE-AC02-07CH11359.
	These resources included support for the \minerva~construction project, and support
	for construction also
	was granted by the United States National Science Foundation under
	Award No. PHY-0619727 and by the University of Rochester. Support for
	participating scientists was provided by NSF and DOE (USA); by CAPES
	and CNPq (Brazil); by CoNaCyT (Mexico); by Proyecto Basal FB 0821, CONICYT PIA ACT1413, Fondecyt 3170845 and 11130133 (Chile); 
	by CONCYTEC (Consejo Nacional de Ciencia, Tecnolog\'{i}a e Innovaci\'{o}n Tecnol\'{o}gica), DGI-PUCP (Direcci\'{o}n de Gesti\'{o}n de la Investigaci\'{o}n  - Pontificia Universidad Cat\'{o}lica del Peru), and VRI-UNI (Vice-Rectorate for Research of National University of Engineering) (Peru);
	and by the Latin American Center for Physics (CLAF); NCN Opus Grant No. 2016/21/B/ST2/01092 (Poland); by Science and Technology Facilities Council (UK).  We thank the MINOS Collaboration for use of its near detector data. Finally, we thank the staff of
	Fermilab for support of the beam line, the detector, and computing infrastructure.

	%
	%
	%
	%
	%
	%
	%
	%
	%
	%
	%

\end{acknowledgments}

  \bibliographystyle{apsrev4}
  \bibliography{2DInclusive}

\end{document}